\documentclass[useAMS,usenatbib]{mnras}

\usepackage{graphicx}
\usepackage{graphics}
\usepackage{color}
\usepackage{url}
\usepackage[caption=false]{subfig}
\usepackage{verbatim}
\usepackage{array}
\usepackage{amssymb}
\usepackage{amsmath}
\usepackage{afterpage}
\usepackage{alphalph}

\usepackage{lscape}
\usepackage{pbox}
\usepackage{supertabular}

\newcommand{\swift}{{\it Swift}}

\volume{484}
\pagerange{1899--1911}
\pubyear{2019}

\begin{document}

\title[ASAS-SN Bright SN Catalog 2017]{The ASAS-SN Bright Supernova Catalog -- IV. 2017}

\author[T.~W.-S.~Holoien et al.]{T.~W.-S.~Holoien$^{1}$\thanks{tholoien@carnegiescience.edu}, J.~S.~Brown$^{2,3}$, P.~J.~Vallely$^{2}$, K.~Z.~Stanek$^{2,4}$, 
\newauthor
C.~S.~Kochanek$^{2,4}$, B.~J.~Shappee$^{5}$, J.~L.~Prieto$^{6,7}$, Subo~Dong$^{8}$, J.~Brimacombe$^{9}$,    
\newauthor
D.~W.~Bishop$^{10}$, S.~Bose$^{8}$, J.~F.~Beacom$^{2,4,11}$, D.~Bersier$^{12}$, Ping~Chen$^{8}$, 
\newauthor 
L.~Chomiuk$^{13}$, E.~Falco$^{14}$, S.~Holmbo$^{15}$, T.~Jayasinghe$^{2}$, N.~Morrell$^{16}$, 
\newauthor
 G.~Pojmanski$^{17}$, J.~V.~Shields$^{2}$, J.~Strader$^{13}$, M.~D.~Stritzinger$^{15}$, 
\newauthor
Todd~A.~Thompson$^{2,4}$, P.~R.~Wo\'zniak$^{18}$, G.~Bock$^{19}$, P.~Cacella$^{20}$, J.~G.~Carballo$^{21}$, 
\newauthor
I.~Cruz$^{23}$, E.~Conseil$^{22}$, R.~G.~Farfan$^{24}$, J.~M.~Fernandez$^{25}$, S.~Kiyota$^{26}$,  
\newauthor
R.~A.~Koff$^{27}$, G. Krannich$^{28}$, P.~Marples$^{29}$, G.~Masi$^{30}$, L.~A.~G.~Monard$^{31}$, 
\newauthor
J.~A.~Mu\~{n}oz$^{32,33}$, B.~Nicholls$^{34}$, R.~S.~Post$^{35}$, G.~Stone$^{36}$, 
\newauthor
D.~L.~Trappett$^{37}$, and W.~S.~Wiethoff$^{38}$\\ \\
  $^{1}$ The Observatories of the Carnegie Institution for Science, 813 Santa Barbara Street, Pasadena, CA 91101, USA \\
  $^{2}$ Department of Astronomy, The Ohio State University, 140 West 18th Avenue, Columbus, OH 43210, USA \\
  $^{3}$ Department of Astronomy and Astrophysics, University of California, Santa Cruz, CA 92064, USA \\
  $^{4}$ Center for Cosmology and AstroParticle Physics (CCAPP), The Ohio State University, 191 W. Woodruff Ave., \\
             \hspace{0.6cm}Columbus, OH 43210, USA \\
  $^{5}$ Institute for Astronomy, University of Hawai'i, 2680 Woodlawn Drive, Honolulu, HI 96822, USA \\
  $^{6}$ N\'ucleo de Astronom\'ia de la Facultad de Ingenier\'ia y Ciencias, Universidad Diego Portales, Av. Ej\'ercito 441, Santiago, Chile \\
  $^{7}$ Millennium Institute of Astrophysics, Santiago, Chile \\
  $^{8}$ Kavli Institute for Astronomy and Astrophysics, Peking University, Yi He Yuan Road 5, Hai Dian District, \\
               \hspace{0.6cm}Beijing 100871, China \\
  $^{9}$ Coral Towers Observatory, Cairns, Queensland 4870, Australia \\
  $^{10}$ Rochester Academy of Science, 1194 West Avenue, Hilton, NY 14468, USA \\
  $^{11}$ Department of Physics, The Ohio State University, 191 W. Woodruff Ave., Columbus, OH 43210, USA \\
  $^{12}$ Astrophysics Research Institute, Liverpool John Moores University, 146 Brownlow Hill, Liverpool L3 5RF, UK \\
  $^{13}$ Department of Physics and Astronomy, Michigan State University, East Lansing, MI 48824, USA \\
  $^{14}$ Harvard-Smithsonian Center for Astrophysics, 60 Garden St., Cambridge, MA 02138, USA \\
  $^{15}$ Department of Physics and Astronomy, Aarhus University, Ny Munkegade 120, DK-8000 Aarhus C, Denmark \\
  $^{16}$ Las Campanas Observatory, Carnegie Observatories, Casilla 601, La Serena, Chile \\
  $^{17}$ Warsaw University Astronomical Observatory, Al. Ujazdowskie 4, 00-478 Warsaw, Poland \\
  $^{18}$ Los Alamos National Laboratory, Mail Stop B244, Los Alamos, NM 87545, USA \\
  $^{19}$ Runaway Bay Observatory, 1 Lee Road, Runaway Bay, Queensland 4216, Australia \\
  $^{20}$ DogsHeaven Observatory, SMPW Q25 CJ1 LT10B, Brasilia, DF 71745-501, Brazil \\
  $^{21}$ Observatorio Cerro del Viento-MPC I84, Paseo Condes de Barcelona, 6-4D, 06010 Badajoz, Spain \\
  $^{22}$ Association Francaise des Observateurs d'Etoiles Variables (AFOEV), Observatoire de Strasbourg, 11 Rue de l'Universite, \\
               \hspace{0.6cm}67000 Strasbourg, France \\
  $^{23}$ Cruz Observatory, 1971 Haverton Drive, Reynoldsburg, OH 43068, USA \\
  $^{24}$ Uraniborg Observatory, Écija, Sevilla, Spain \\
  $^{25}$ Observatory Inmaculada del Molino, Hernando de Esturmio 46, Osuna, 41640 Sevilla, Spain \\
  $^{26}$ Variable Star Observers League in Japan, 7-1 Kitahatsutomi, Kamagaya, Chiba 273-0126, Japan \\
  $^{27}$ Antelope Hills Observatory, 980 Antelope Drive West, Bennett, CO 80102, USA \\
  $^{28}$ Roof Observatory Kaufering, Lessingstr. 16, D-86916 Kaufering, Germany \\
  $^{29}$ Leyburn \& Loganholme Observatories, 45 Kiewa Drive, Loganholme, Queensland 4129, Australia \\
  $^{30}$ Virtual Telescope Project, Via Madonna de Loco, 47-03023 Ceccano (FR), Italy \\
  $^{31}$ Kleinkaroo Observatory, Calitzdorp, St. Helena 1B, P.O. Box 281, 6660 Calitzdorp, Western Cape, South Africa \\
  $^{32}$ Departamento de Astronom\'{\i}a y Astrof\'{\i}sica, Universidad de Valencia, E-46100 Burjassot, Valencia, Spain \\
  $^{33}$ Observatorio Astron\'omico, Universidad de Valencia, E-46980 Paterna, Valencia, Spain
  $^{34}$ Mount Vernon Observatory, 6 Mount Vernon Place, Nelson, New Zealand \\
  $^{35}$ Post Observatory, Lexington, MA 02421, USA \\
  $^{36}$ Sierra Remote Observatories, 44325 Alder Heights Road, Auberry, CA 93602, USA \\
  $^{37}$ Brisbane Girls Grammar School - Dorothy Hill Observatory, Gregory Terrace, Spring Hill, Queensland 4000, Australia \\
  $^{38}$ Department of Earth and Environmental Sciences, University of Minnesota, 230 Heller Hall, 1114 Kirby Drive, \\
               \hspace{0.6cm}Duluth, MN. 55812, USA
  }
\maketitle
\begin{abstract}
In this catalog we compile information for all supernovae discovered by the All-Sky Automated Survey for SuperNovae (ASAS-SN) as well as all other bright ($m_{peak}\leq17$), spectroscopically confirmed supernovae found in 2017, totaling 308 supernovae. We also present UV through near-IR magnitudes gathered from public databases of all host galaxies for the supernovae in the sample. We perform statistical analyses of our full bright supernova sample, which now contains 949 supernovae discovered since 2014 May 1, including supernovae from our previous catalogs. This is the fourth of a series of yearly papers on bright supernovae and their hosts from the ASAS-SN team, and this work presents updated data and measurements, including light curves, redshifts, classifications, and host galaxy identifications, that supersede information contained in any previous publications.
\end{abstract}

\raggedbottom

%%%%%%%%%%%%%%%%%
% Section: Introduction
%%%%%%%%%%%%%%%%%

\section{Introduction}
\label{sec:intro}

In recent years, large-scale, systematic surveys for supernovae (SNe) and other transient phenomena have become a cornerstone of modern astronomy. (Significant examples of such surveys are summarized in Paper III; \citealt{holoien17b}). Despite the large number of transient surveys, however, prior to 2013 there was no high-cadence optical survey designed to scan the full observable sky for the bright, nearby transients that are easiest to observe in great detail. While fewer in number, such events provide the opportunity to obtain the high-quality observational data needed to have the largest impact on our understanding of the physics behind transient phenomena.

The All-Sky Automated Survey for SuperNovae (ASAS-SN\footnote{\url{http://www.astronomy.ohio-state.edu/~assassin/}}; \citealt{shappee14}) was created for this purpose. ASAS-SN is designed to survey the entire visible sky at a rapid cadence to find the brightest transients. ASAS-SN has found many nearby and interesting SNe \citep[e.g.,][]{dong16,holoien16c,shappee16,godoy-rivera17,bose18a,bose18b,vallely19}, tidal disruption events \citep[TDEs; e.g.,][]{holoien14b,brown16a,brown17a,holoien16b,holoien16a,prieto16,romero16,holoien18a}, stellar outbursts \citep{holoien14a,schmidt14,herczeg16,schmidt16}, flares from active galactic nuclei \citep{shappee14}, black hole binaries \citep{tucker18}, and cataclysmic variable stars \citep{kato13,kato14,kato15,kato16}.

Each ASAS-SN unit is hosted by the Las Cumbres Observatory \citep{brown13} network and consists of four 14-cm telescopes, each with a $4.5\times4.5$ degree field-of-view. Until 2017, ASAS-SN comprised two units, each using $V$-band filters with a limiting magnitude of $m_V\sim17$: Brutus, located on Haleakala in Hawaii, and Cassius, located at Cerro Tololo, Chile (further technical details can be found in \citet{shappee14}). In late 2017, ASAS-SN expanded with three new units: Paczynski, also located at Cerro Tololo, Leavitt, located at McDonald Observatory in Texas, and Payne-Gaposchkin, located in Sutherland, South Africa. Between the five units, ASAS-SN now surveys the entire visible sky (roughly 30000 square degrees) in less than a single night, with weather losses. Further, ASAS-SN switched to $g$-band for our new units, increasing our imiting magnitude to $m_g\sim18.5$. Due to the increased depth, we will be switching our initial 8 telescopes to $g$-band as well by the end of 2018. For a more detailed history of the ASAS-SN project, see \citet{holoien16d} and \citet{shappee14}.

All ASAS-SN data are automatically processed and are searched in real-time. New discoveries are announced publicly either upon first detection if the source is judged likely to be real based on ASAS-SN data alone, or once confirmed through follow-up imaging in cases where the first detection is ambiguous. This allows for both rapid discovery and response by the ASAS-SN team, as well as by others. ASAS-SN uses an untargeted survey approach and 97\% of announced ASAS-SN discoveries have been confirmed spectroscopically, either by the ASAS-SN team or by public classification efforts via channels such as the Transient Name Server (TNS\footnote{\url{https://wis-tns.weizmann.ac.il/}}) and Astronomer's Telegrams (ATels). This makes the ASAS-SN sample much less biased than many other SN searches, and it is thus ideal for population studies of nearby SNe and their hosts \citep[e.g.,][]{brown18}.

This manuscript is the fourth of a series of yearly catalogs provided by the ASAS-SN team. We present collected information on all SNe discovered by ASAS-SN in 2017 and their host galaxies. We also provide the same information for all bright SNe (those with $m_{peak}\leq17$) discovered by other professional and amateur astronomers in the same year, as we have done with our previous catalogs \citep{holoien16d,holoien17a,holoien17b}. We also include whether ASAS-SN independently recovered these SNe after their initial discovery, to better quantify the completeness of our survey.

This paper contains updated measurements and analysis that are intended to supersede publicly available information from ASAS-SN webpages, TNS, and discovery and classification ATels. ASAS-SN continues to participate in the TNS system to avoid potential confusion over discoveries, but strongly object to a naming scheme that does not credit the discoverer, even though this would be trivial to implement. Throughout this catalog we use the internal survey discovery names for each SN as our primary nomenclature, and we encourage others to do the same to preserve the origins of new transients in future literature.

The catalog is organized as follows: in Section~\ref{sec:sample} we describe the sources of the information presented in this manuscript and list SNe with updated classifications and redshift measurements. In Section~\ref{sec:analysis}, we give statistics on the SN and host galaxy populations in our full cumulative sample, including the discoveries listed in \citet{holoien16d,holoien17a,holoien17b}, and discuss overall trends in the sample. Throughout our analyses, we assume a standard $\Lambda$CDM cosmology with $H_0=69.3$~km~s$^{-1}$~Mpc$^{-1}$, $\Omega_M=0.29$, and $\Omega_{\Lambda}=0.71$ for converting host redshifts into distances. In Section~\ref{sec:disc}, we summarize our overall findings and discuss future directions for the ASAS-SN survey and how they will impact future discoveries.

%%%%%%%%%%%%%%%%%
% Section: Sample
%%%%%%%%%%%%%%%%%

\section{Data Samples}
\label{sec:sample}

The sources of the data collected in our supernova and galaxy samples are summarized below. All data are presented in  Tables~\ref{table:asassn_sne}, \ref{table:other_sne}, \ref{table:asassn_hosts}, and \ref{table:other_hosts}.

%%%%%%%%%%%%%%%%%
% Subsection: ASAS-SN Sample
%%%%%%%%%%%%%%%%%

\subsection{The ASAS-SN Supernova Sample}
\label{sec:asassn_sample}

All information for supernovae discovered by ASAS-SN between 2017 January 1 and 2017 December 31 is given in Table~\ref{table:asassn_sne}. As in \citet{holoien16d,holoien17a,holoien17b}, we obtained all supernova names, discovery dates, and host names from our discovery ATels, which are cited in Table~\ref{table:asassn_sne}. Also included in the table are the IAU names given to each supernova by TNS, which is the official mechanism for reporting new astronomical transients to the IAU. 

We measured all ASAS-SN supernova redshifts from classification spectra. For cases where the nominal supernova host galaxy had a previously measured redshift that is consistent with the transient redshift, we list the redshift of the host taken from the NASA/IPAC Extragalactic Database (NED)\footnote{\url{https://ned.ipac.caltech.edu/}}. For other cases, we report the redshifts given in the classification telegrams or posted on TNS, with the exception of those that are updated in this work (see below).

Classifications for ASAS-SN supernova discoveries were retrieved from classification telegrams, which are cited in Table~\ref{table:asassn_sne} when available, or from TNS, when a classifcation was not reported in an ATel. We list ``TNS'' in the ``Classification Telegram'' column of the table for such cases. When measurable from the classification spectra, we also give the approximate age of the supernova at discovery in days relative to peak. Classifications were typically obtained using either the Supernova Identification code \citep[SNID;][]{blondin07} or the Generic Classification Tool (GELATO\footnote{\url{gelato.tng.iac.es}}; \citealt{harutyunyan08}), which both compare observed input spectra to template spectra in order to estimate the supernova age and type.

We present updated redshifts and classifications for a number of ASAS-SN discoveries whose redshifts and classifications differ from previous reports based on an examination of archival classification spectra obtained from TNS and the Weizmann Interactive Supernova data REPository \citep[WISEREP;][]{yaron12}. ASASSN-17bb, ASASSN-17ol, and ASASSN-17om have updated redshifts, and ASASSN-17io has an updated type based on archival spectra. We also report the classifications (and in some cases, redshifts) for a number of supernovae that were not previously publicly classified based on spectra obtained with the Wide Field Reimaging CCD Camera (WFCCD) mounted on the Las Campanas Observatory du Pont 2.5-m telescope and the Fast Spectrograph \citep[FAST;][]{fabricant98} mounted on the Fred L. Whipple Observatory Tillinghast 1.5-m telescope. ASAS-SN discoveries with new classifications include ASASSN-17ip, ASASSN-17mf, ASASSN-17oh, and ASASSN-17ot. All updated redshifts and classifications are available in Table~\ref{table:asassn_sne}.

We used the astrometry.net package \citep{barron08,lang10} to solve astrometry in follow-up images of ASAS-SN supernovae and used {\sc Iraf} to measure centroid positions for each SN. This generally gives position errors of $<$1\farcs{0} and is substantially more accurate than using ASAS-SN images, which have a 7\farcs{0} pixel scale, to measure SN positions. Follow-up imaging used to measure positions were obtained using the Las Cumbres Observatory 1-m telescopes or by amateur collaborators who work with the ASAS-SN team. Coordinates measured from follow-up imaging were announced in discovery ATels, and we report these coordinates in Table~\ref{table:asassn_sne}. The offsets between the SNe and the centers of their host galaxies are also reported in the Table, and these offsets were calculated using galaxy coordinates available in NED, or measured from archival images in cases where a host center was not previously catalogued.

We re-processed ASAS-SN data and measured $V$-band, host-subtracted light curves for all ASAS-SN supernova discoveries spanning 90 days prior to discovery through 250 days after discovery. From these light curves we measured $V$-band peak magnitudes for each ASAS-SN supernova. In addition, as the three new units deployed in 2017 use $g$-band filters, we measured $g$-band peak magnitudes for each ASAS-SN supernova that was discovered after the new units were deployed. Both magnitudes are reported in Table~\ref{table:asassn_sne}. In some cases, due to the way ASAS-SN fields were divided between cameras and because the new units were still building reference images in late 2017, supernovae were only detected in a single filter. We also report discovery magnitudes in the discovery filter measured from the re-subtracted light curves. In some cases, these magnitudes differ from the magnitudes reported in the original discovery ATels or on TNS, as re-processing the data can result in improvements in the photometry. As we did in the previous ASAS-SN catalogs, we define the ``discovery magnitude'' as the supernova's magnitude on the date of discovery. For supernovae with enough detections in their light curves (for either or both filters), we also performed parabolic fits to the full light curves and estimate peak magnitudes based on the fits. We report the brighter value between the brightest magnitude measured from the light curve and the peak of the parabolic fit for each filter as the ``peak magnitudes'' in Table~\ref{table:asassn_sne}This is done to provide more accurate peak magnitudes for the few cases where the peak of the light curve was not well-covered by ASAS-SN data, as was done in the previous ASAS-SN catalogs.

As in \citet{holoien16d,holoien17a,holoien17b}, all supernovae discovered by ASAS-SN in 2017 are included in this catalog, including those fainter than $m_V=17$ or $m_g=17$. When performing comparison analyses that are presented in Section~\ref{sec:analysis}, we only include those ASAS-SN discoveries with $m_{peak}\leq17$ so that our sample is consistent with the non-ASAS-SN sample.

%%%%%%%%%%%%%%%%%
% Subsection: Other SN Sample
%%%%%%%%%%%%%%%%%

\subsection{The Non-ASAS-SN Supernova Sample}
\label{sec:other_sample}

Table~\ref{table:other_sne} contains information for all SNe discovered by other professional and amateur SN searches between 2017 January 1 and 2017 December 31 that are both spectroscopically confirmed and have peak magnitudes of $m_{peak}\leq17$.

%%%%%%%%%%%%%%%%%
% Figure: Pie Charts
%%%%%%%%%%%%%%%%%

\begin{figure*}
\begin{minipage}{\textwidth}
\centering
\subfloat{{\includegraphics[width=0.31\textwidth]{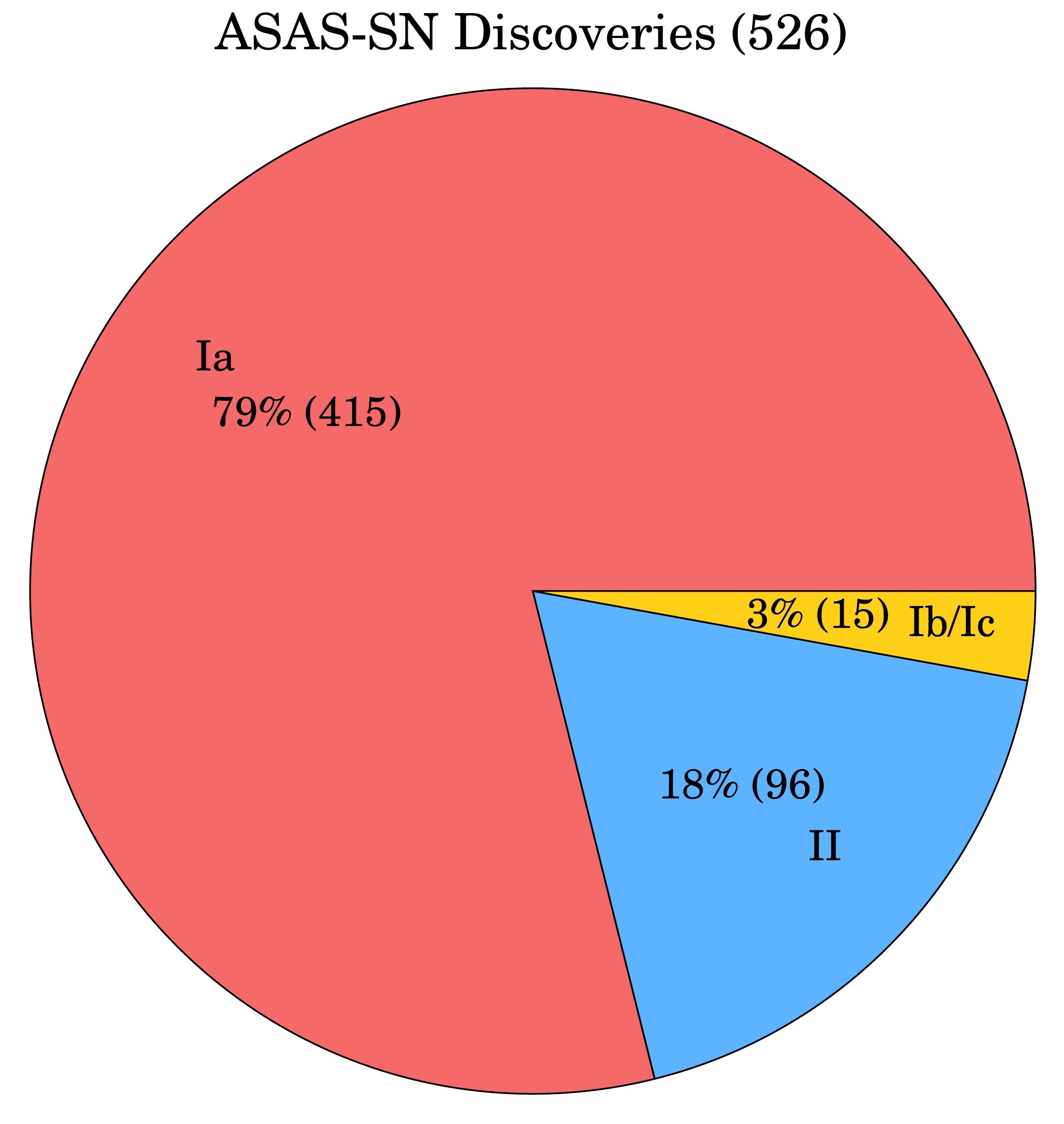}}}
\subfloat{{\includegraphics[width=0.31\textwidth]{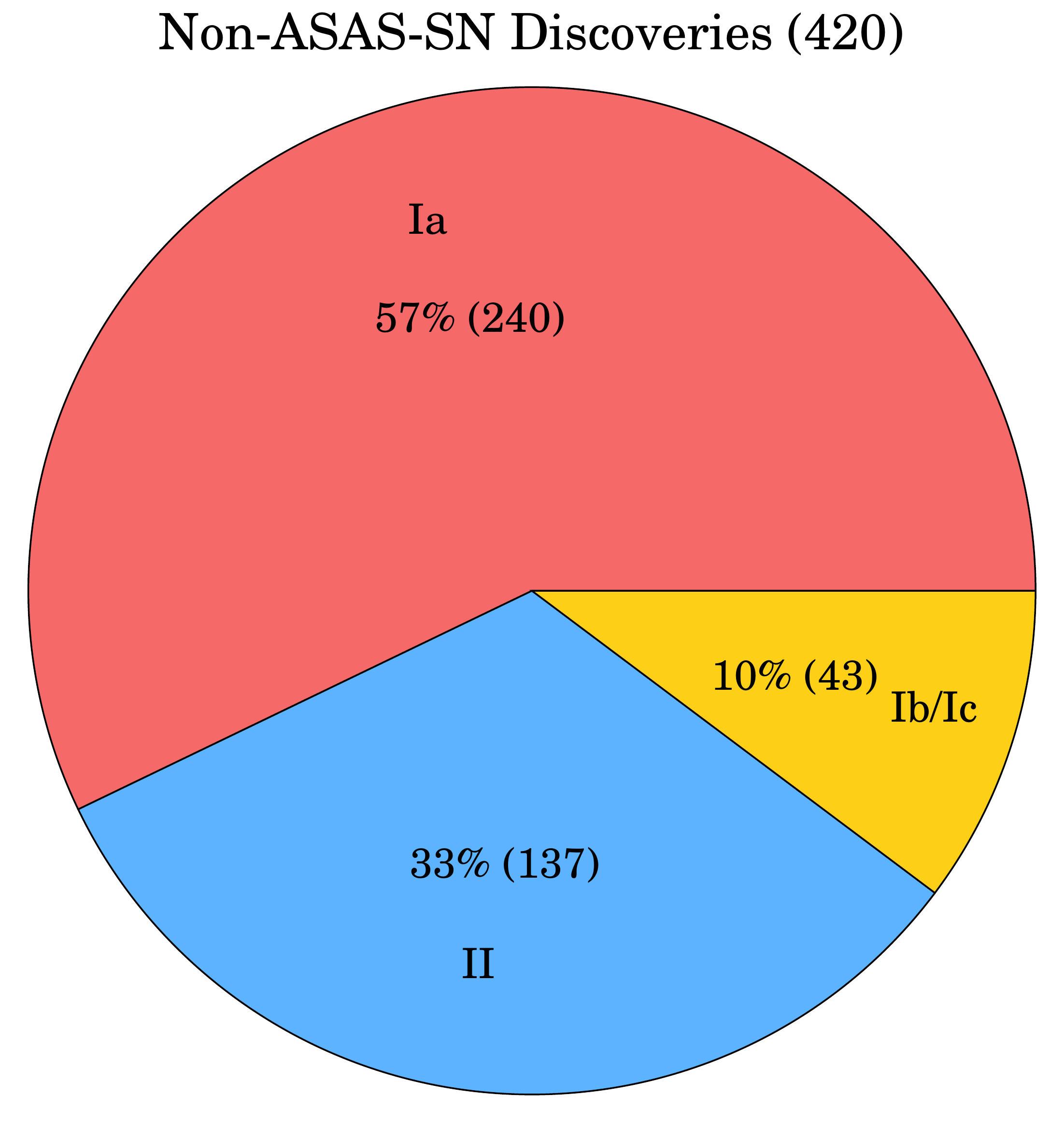}}}
\subfloat{{\includegraphics[width=0.31\textwidth]{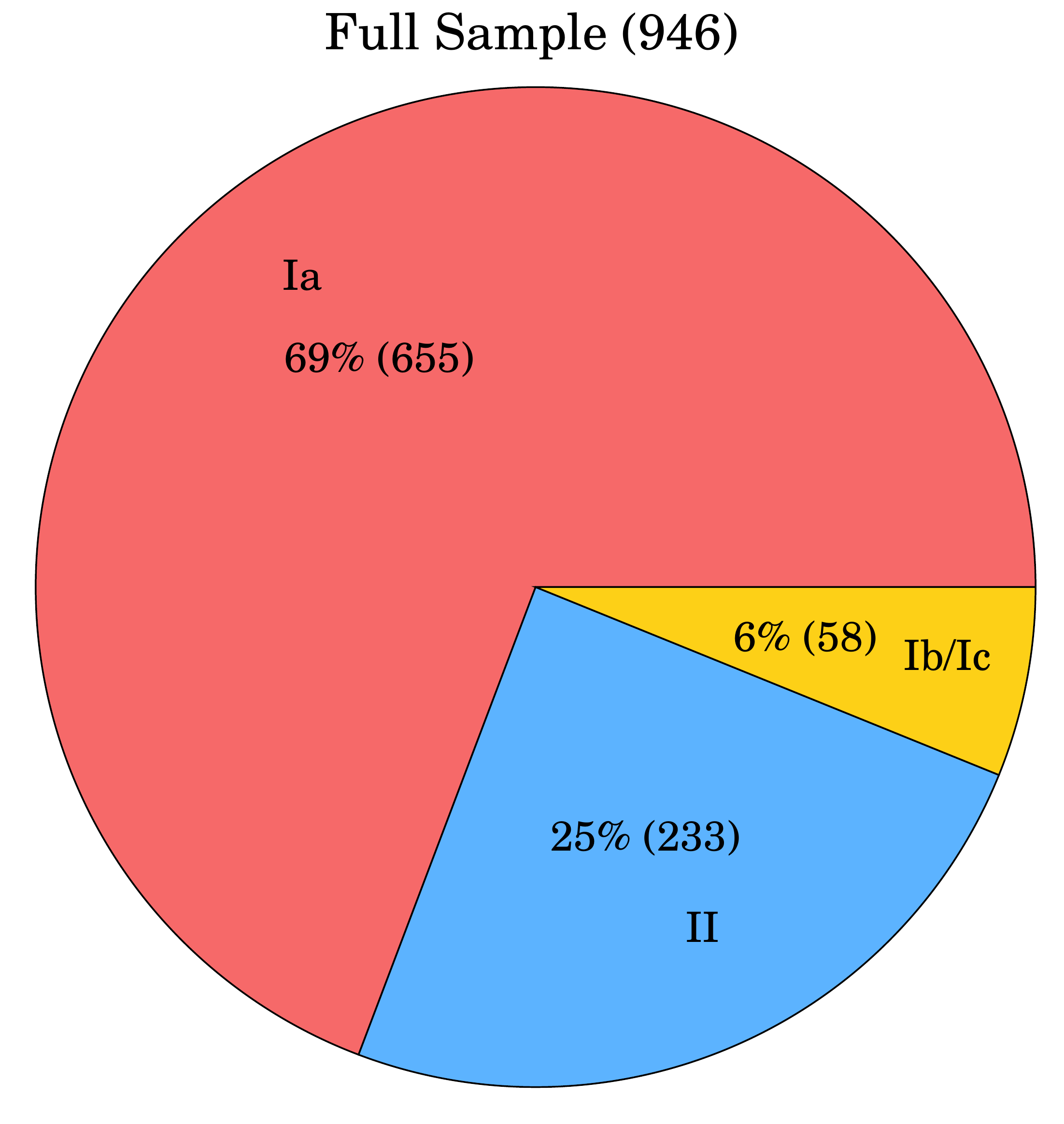}}}
\caption{\emph{Left Panel}: Breakdown by type of ASAS-SN supernova discoveries between 2014 May 01 and 2017 December 31. The proportion of each type continues to closely match that of the ideal magnitude-limited sample from \citet{li11}. \emph{Center Panel}: The same breakdown for the non-ASAS-SN sample over the the same time period. \emph{Right Panel}: The type breakdown for our entire SN sample, now totaling 946 supernovae.} This analysis excludes the 3 superluminous SNe in the sample, and includes Type IIb SNe in the ``Type II'' sample.
\label{fig:piechart}
\end{minipage}
\end{figure*}

As in our previous catalogs, the list of non-ASAS-SN discoveries was gathered from the ``latest supernovae'' website\footnote{\url{http://www.rochesterastronomy.org/snimages/}} created by D.~W.~Bishop \citep{galyam13}. This website compiles discoveries reported via a multitude of channels, including TNS and ATels, and links objects reported by different searches at different times. It is thus an ideal source for information on discoveries from various SN searches. Some supernova searches are not participating in the TNS system, so we only use TNS for verifying data on the latest supernovae website, and not as a primary information source.

We obtained supernova names, IAU names, discovery dates, coordinates, host offsets, peak reported magnitudes, spectral types, and discovery sources for each supernovae in the non-ASAS-SN sample from the latest supernovae website, when possible. NED was used to gather host galaxy names and redshifts when available, with the SN redshifts on the latest supernovae website being used in other cases. For supernovae without a host offset listed on the website, we took the offset from NED, defining the offset as the angular separation between the coordinates of the host in NED and the reported SN coordinates.

Some SNe has no host galaxy listed near their positions in NED, but had hosts clearly visible in archival Pan-STARRS data \citep{chambers16}. For such cases we used {\sc Iraf} to measure a centroid position of the host to use to calculate the offset. The host galaxy names given in this manuscript are the primary names of the galaxies taken from NED. These differ in some cases from what is listed on ASAS-SN websites or the latest supernovae website.

The magnitudes from the latest supernovae website are reported in different filters from various telescopes, and in many cases the reported photometry does not necessarily cover the actual peak of the supernova light curve. For the purposes of having a more consistent sample of supernova peak magnitudes between the ASAS-SN sample, which uses peak magnitudes measured from ASAS-SN data, and the non-ASAS-SN sample, we also produced new, host-subtracted $V$- and $g$-band ASAS-SN light curves for every non-ASAS-SN supernova in the 2017 sample. As we did with the ASAS-SN sample, we also performed parabolic fits to the light curves with enough detections, and used the brighter of the brightest measured magnitude and the peak of the fit as the ``peak magnitude'' for each filter, when a supernova was detected. These peak ASAS-SN $V$- and $g$-band magnitudes are also listed in Table~\ref{table:other_sne} for each supernova that was detected. We find that only 9 supernovae from the 2017 non-ASAS-SN sample are not detected in this re-examination despite 48 of these supernovae not being recovered during normal survey operations. This is likely due to better-quality light curves being produced in this new reduction, and the fact that our processing ensures no supernova light is contained in the reference image, and thus subtracted from the light curve.

We performed a similar reduction and peak magnitude measurement for supernovae in the 2014-2016 non-ASAS-SN samples, and use only ASAS-SN $V$-band magnitudes when looking at the peak magnitude distribution and sample completeness in Section~\ref{sec:analysis}. The ASAS-SN light curves for all supernovae in these samples will be released in a future manuscript (Ping et al., \emph{in prep.})

As we did with the ASAS-SN sample, we also checked the redshifts and classifications of the non-ASAS-SN supernovae using publicly available spectra on TNS and WISEREP. Based on our re-examination of these spectra, we update the classifications of ATLAS17cpj and ATLAS17evm and the redshift of ATLAS17cpj. In addition, the supernova SN 2017gfj has a measured redshift of $z\sim0.072$, but has been publicly announced as being hosted in the galaxy UGC 11950, which has a redshift of $z=0.020541$. This was likely done because UGC 11950 is the nearest catalogued galaxy to the SN, but based on the redshift discrepancy we believe it likely that SN 2017gfj was actually located in an uncatalogued background galaxy, and we update the host name in our sample accordingly. We assume the SN redshift of $z\sim0.072$ for SN 2017gfj in our analyses presented in Section~\ref{sec:analysis}. We report updated types and redshifts in Table~\ref{table:other_sne}.

We give the name of the discovery group for all SNe discovered by other professional surveys in Table~\ref{table:other_sne}. We use ``Amateurs'' as the discovery source for supernovae discovered by non-professional astronomers, as this allows us to differentiate this sample of SNe from the ASAS-SN and other professional samples. We include the names of the amateurs responsible for these discoveries in the full machine-readable version of Table~\ref{table:other_sne} that is available in the online version of this manuscript so as to properly credit them for their discoveries. Unlike in previous years, amateurs no longer account for the second largest number of bright supernova discoveries after ASAS-SN, with the ATLAS survey \citep{tonry11,tonry18} now holding that distinction. Amateurs still account for the third largest number of bright supernova discoveries in 2017, however, showing they are still a significant source of bright discoveries.

Finally, we also note in Table~\ref{table:other_sne} whether or not non-ASAS-SN supernovae were independently recovered by the ASAS-SN team during normal survey operations. This allows us to better quantify the impact of ASAS-SN on the discovery of bright SNe independent of other SN searches.

%%%%%%%%%%%%%%%%%
% Subsection: Host Galaxy Sample
%%%%%%%%%%%%%%%%%

\subsection{The Host Galaxy Samples}
\label{sec:host_sample}

For both the ASAS-SN and non-ASAS-SN samples, we collected Galactic extinction estimates in the directions of the host galaxies and near-ultraviolet (NUV) through infrared (IR) host galaxy magnitudes, which we presentin Tables~\ref{table:asassn_hosts} and \ref{table:other_hosts}. We obtained the values of Galactic $A_V$ from \citet{schlafly11} in the directions of the supernovae from NED. NUV host magnitudes from the Galaxy Evolution Explorer \citep[GALEX;][]{morrissey07} All Sky Imaging Survey (AIS), optical $ugriz$ magnitudes from the Sloan Digital Sky Survey Data Release 14 \citep[SDSS DR14;][]{albareti16}, NIR $JHK_S$ magnitudes from the Two-Micron All Sky Survey \citep[2MASS;][]{skrutskie06}, and IR $W1$ and $W2$ from the Wide-field Infrared Survey Explorer \citep[WISE;][]{wright10} AllWISE source catalogs were obtained from publicly available online databases. 

%%%%%%%%%%%%%%%%%
% Figure: Offset-Mag
%%%%%%%%%%%%%%%%%

\begin{figure*}
\begin{minipage}{\textwidth}
\centering
\subfloat{{\includegraphics[width=0.85\textwidth]{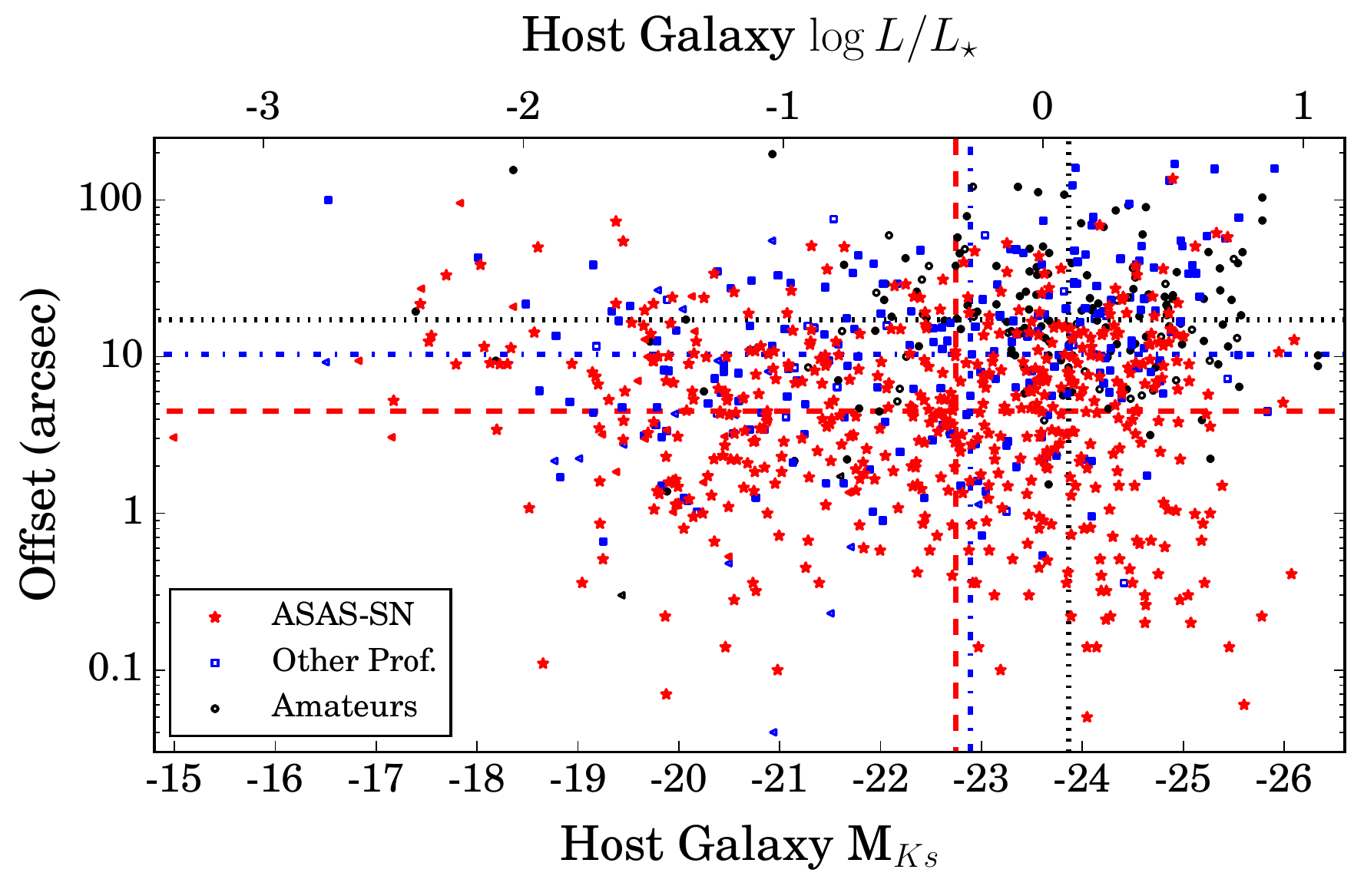}}}

\centering
\subfloat{{\includegraphics[width=0.85\textwidth]{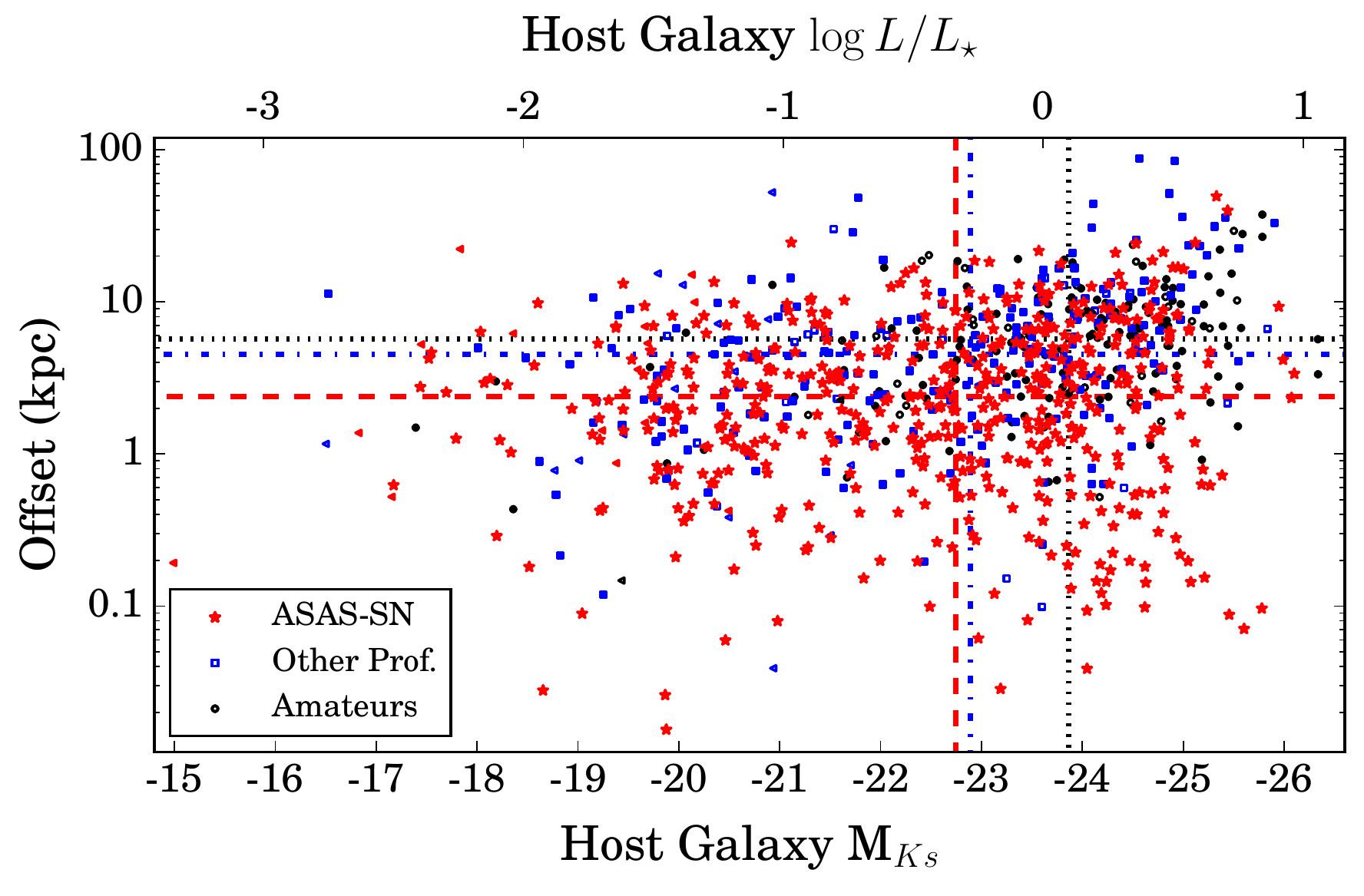}}}
\caption{\emph{Upper Panel}: The absolute $K_S$ host magnitudes for all supernovae in our $2014-2017$ combined sample compared to the offsets of the SNe in arcseconds from their host nuclei. The top scale shows a luminosity scale corresponding to the magnitude range on the bottom scale, assuming $M_{\star,K_S}=-24.2$ \citep{kochanek01}. Red stars, black circles, and blue squares denote discoveries from ASAS-SN, amateurs, and other professional searches, respectively. Triangles in corresponding colors indicate upper limits on the host galaxy magnitudes for hosts that are not detected in 2MASS or WISE data. Filled points indicate supernovae that were discovered or independently recovered by ASAS-SN. The median host magnitudes and offsets are indicated with dashed, dotted, and dash-dotted lines for the ASAS-SN sample, amateur sample, and other professional sample, respectively, in colors that correspond to those of the matching data points. \emph{Lower Panel}: The same plot, but with the offset measured in kiloparsecs.}
\label{fig:offmag}
\end{minipage}
\end{figure*}

We adopt $J$ and $H$ filter upper limits corresponding to ther faintest detected host in our combined $2014-2017$ sample ($m_J>17.0$, $m_H>16.4$) for hosts that are not detected in 2MASS data. If a host not detected in 2MASS was detected in WISE $W1$ data, we added the mean $K_S-W1$ offset from the total sample to the WISE $W1$ magnitude to estimate a $K_S$ magnitude. This average offset is equal to $-0.51$ magnitudes with a scatter of $0.04$ magnitudes and a standard error of $0.002$ magnitudes, matching what we found when doing the same calculation for the $2014-2016$ sample in \citet{holoien17b}. We adopted an upper limit of $m_{K_S}>15.6$, equal to the faintest detected $K_S$-band host magnitude from our sample, for galaxies not detected in either 2MASS or WISE data.

%%%%%%%%%%%%%%%%%
% Figure: Offset-Mag Distributions
%%%%%%%%%%%%%%%%%

\begin{figure*}
\begin{minipage}{\textwidth}
\centering
\subfloat{{\includegraphics[width=0.75\textwidth]{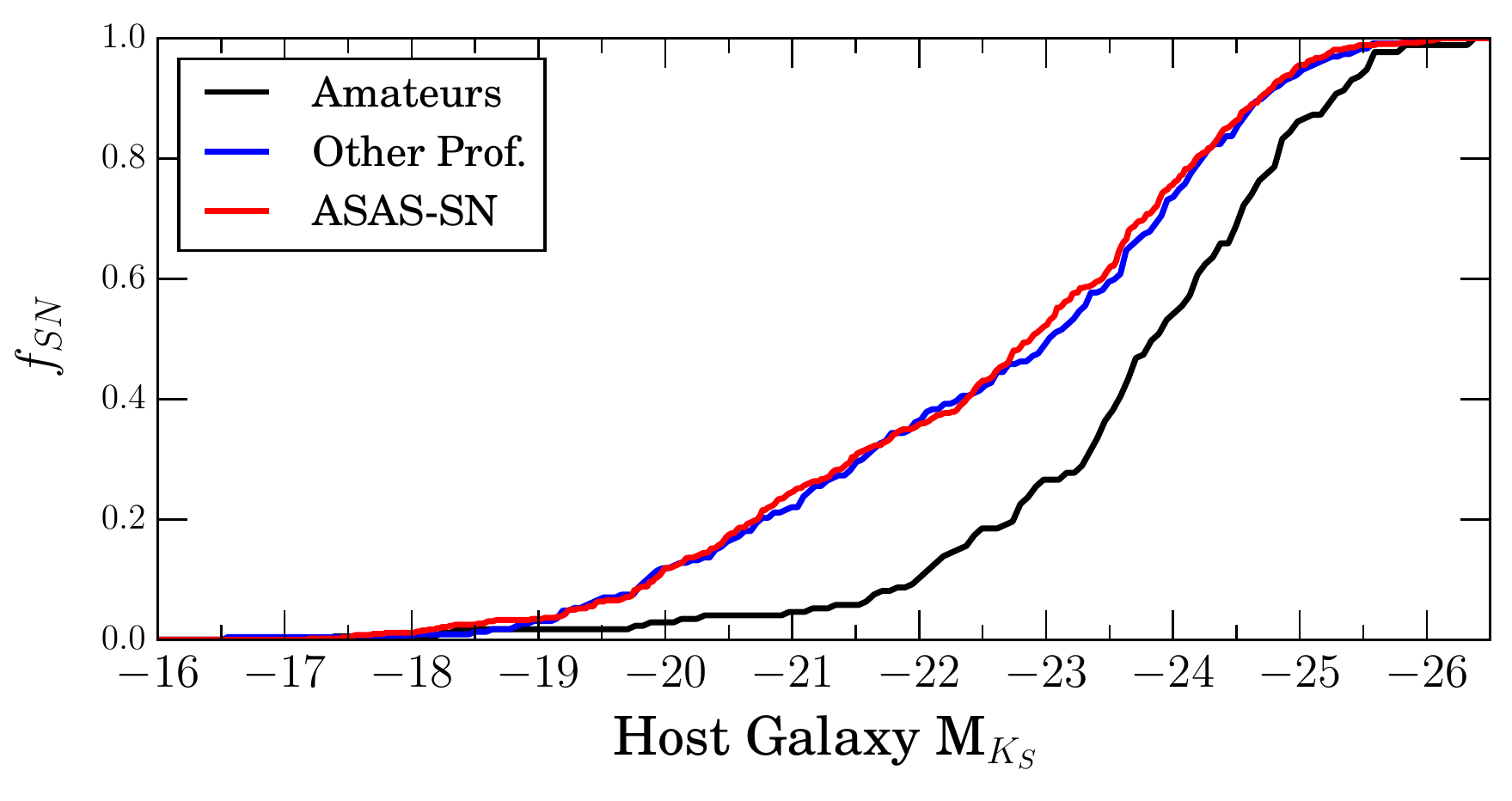}}}

\centering
\subfloat{{\includegraphics[width=0.75\textwidth]{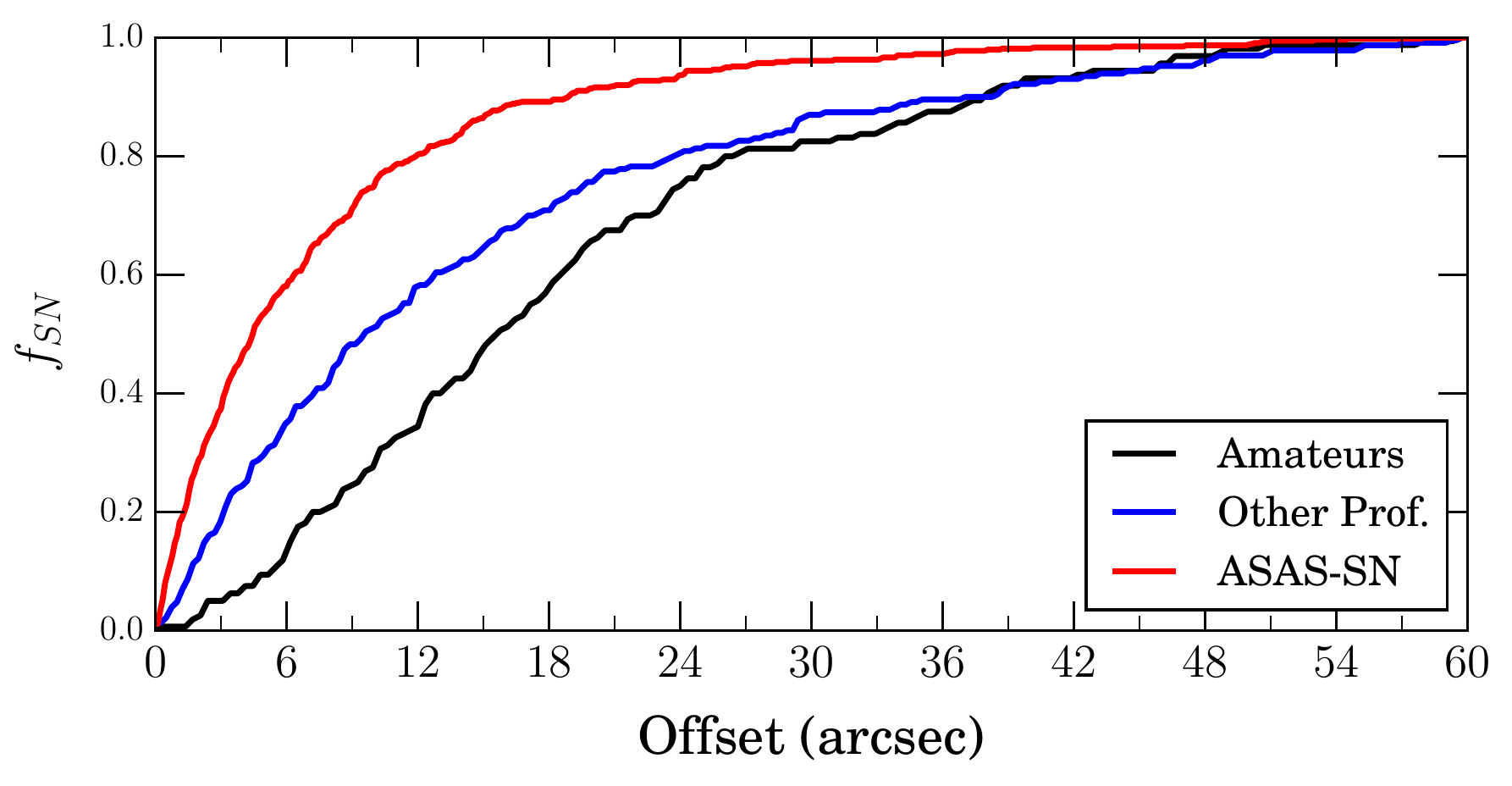}}}

\centering
\subfloat{{\includegraphics[width=0.75\textwidth]{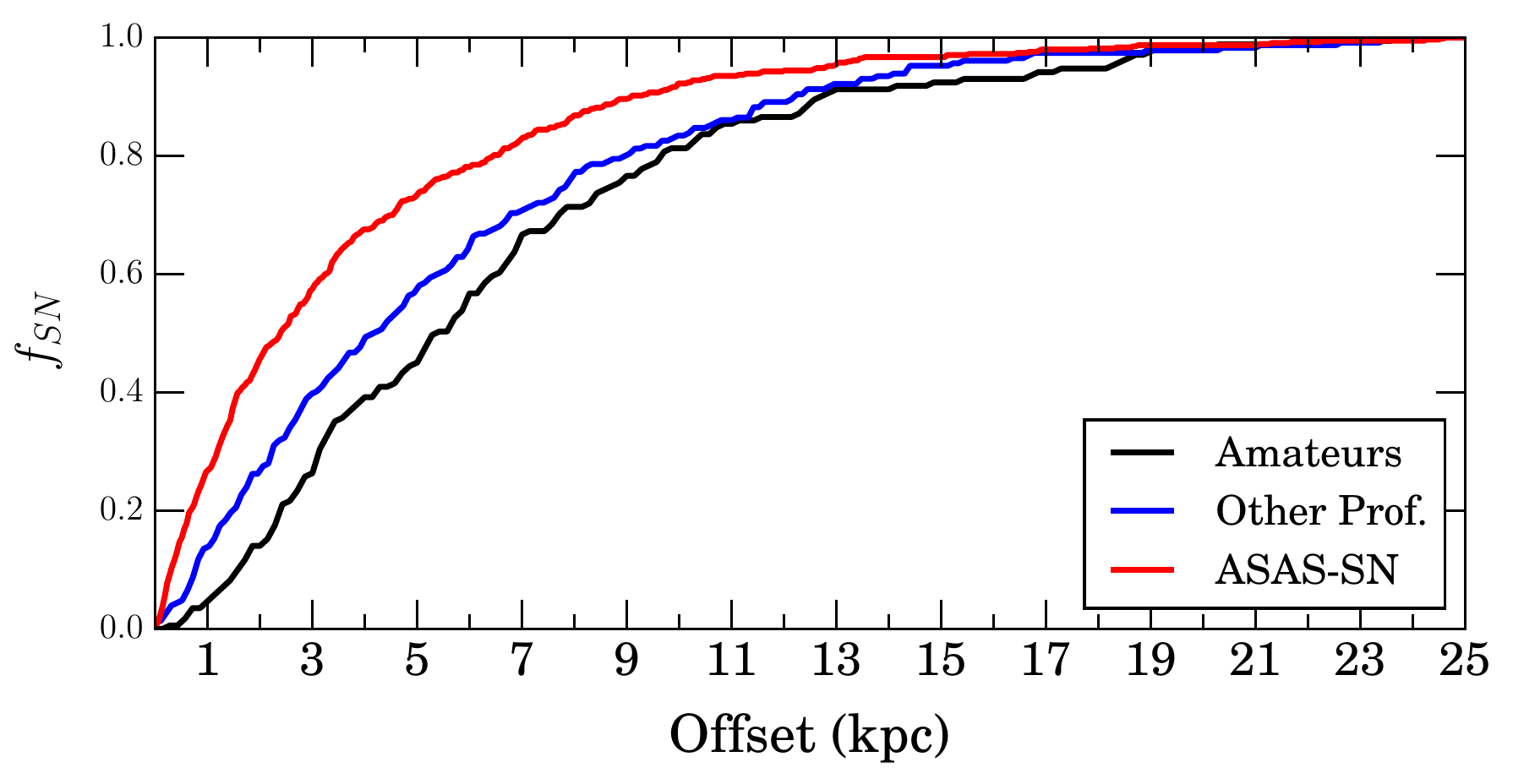}}}
\caption{Normalized cumulative distributions of host galaxy absolute magnitudes (upper panel), SN offsets from host nuclei in arcseconds (center panel), and SN offsets from host nuclei in kpc (bottom panel) for ASAS-SN discoveries (red), other professional discoveries (blue), and amateur discoveries (black). As is seen in Figure~\ref{fig:offmag}, amateur discoveries are biased towards more luminous hosts than professional surveys, and ASAS-SN continues to find supernovae at smaller offsets than either comparison group, regardless of how the offset is measured.}
\label{fig:offmag_dist}
\end{minipage}
\end{figure*}

%%%%%%%%%%%%%%%%%
% Section: Analysis
%%%%%%%%%%%%%%%%%

\section{Analysis of the Sample}
\label{sec:analysis}

The total sample of bright SNe discovered by all sources between 2014 May 01, when ASAS-SN began operations in the Southern hemisphere, and 2017 December 31 now includes 949 SNe, after excluding ASAS-SN discoveries with $m_{peak,V}>17.0$ and $m_{peak,g}>17.0$ \citep{holoien16d,holoien17a,holoien17b}. 56\% (528) of these SNe were ASAS-SN discoveries, 19\% (176) were discovered by amateur astronomers, and 26\% (245) were discovered by other professional surveys. Breaking the sample down by type, 655 were Type Ia SNe, 233 were Type II SNe, 58 were type Ib/Ic SNe, and 3 were superluminous SNe. For the purpose of these analyses, we include Type IIb SNe in the Type II sample so that we can more directly compare with the results of \citet{li11}, as we have done in our previous catalogs. The object ASASSN-15lh, either an extremely luminous Type I SLSN \citep{dong16,godoy-rivera17} or a tidal disruption event \citep{leloudas16}, is excluded from the following analysis.

%%%%%%%%%%%%%%%%%
% Figure: 2017 Histogram
%%%%%%%%%%%%%%%%%

\begin{figure*}
\begin{minipage}{\textwidth}
\centering
\subfloat{{\includegraphics[width=0.95\linewidth]{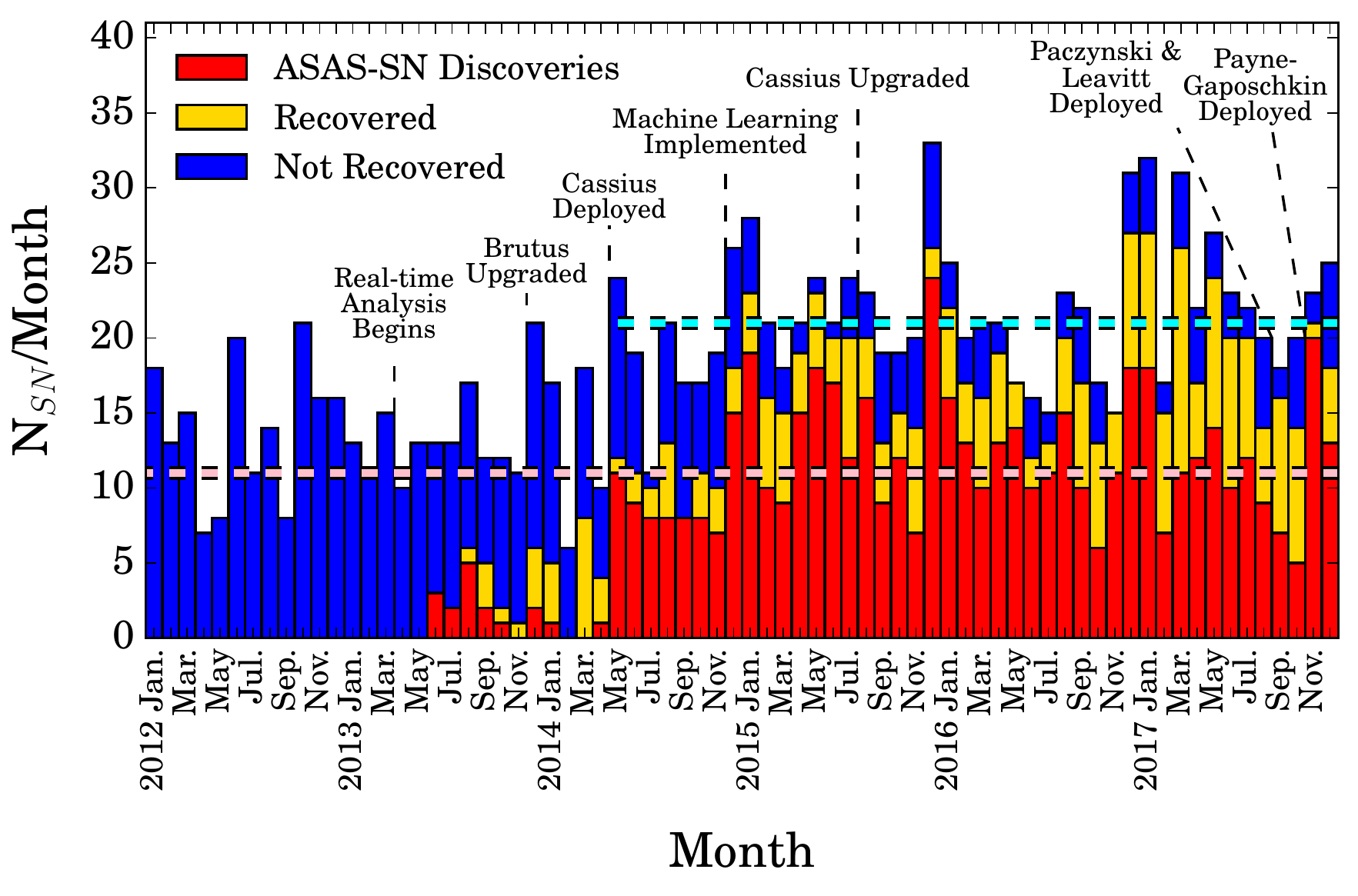}}}
\caption{Histogram of bright supernovae discovered in each month from 2012 through 2017. ASAS-SN discoveries are shown in red, supernovae found by other sources and recovered in ASAS-SN data are shown in yellow, and supernovae not recovered by ASAS-SN (or discovered prior to ASAS-SN coming online) are shown in blue. Significant milestones in the ASAS-SN timeline are indicated. The dashed pink line indicates the median number of bright supernovae discovered in each month from 2010 through 2012, while the dashed cyan line indicates the median number of bright supernovae discovered in each month since 2014 May. Since ASAS-SN became operational in both hemispheres in 2014 May, the number of bright supernova discoveries has exceeded the previous median and supernovae discovered or recovered by ASAS-SN account for more than half of all bright supernova discoveries in every month.}
\label{fig:histogram}
\end{minipage}
\end{figure*}

The breakdown by type of the ASAS-SN, non-ASAS-SN, and total samples are shown in Figure~\ref{fig:piechart}. As \citep[e.g.,][]{li11} predicted for a magnitude-limited sample, Type Ia SNe account for the largest fraction in each of the three samples. As we have seen in our previous catalogs, the ASAS-SN sample continues to match the ``ideal magnitude-limited sample'' from \citet{li11}, where there are 79\% Type Ia, 17\% Type II, and 4\% Type Ib/Ic SNe, almost exactly. Due to the observing strategies of the various discovery sources in the non-ASAS-SN sample not necessarily being magnitude-limited in all cases (e.g., because the survey targets certain types of galaxies or takes a volume-limited approach), the other two samples have higher fractions of core-collapse SNe than the ASAS-SN sample.

ASAS-SN accounts for 56\% of the bright supernovae in our total sample, and thus remains the dominant source of bright supernova discoveries despite new surveys like ATLAS coming online in 2017. A large fraction of the ASAS-SN sample continues to be discovered shortly after explosion because of our high cadence. Out of 459 ASAS-SN SNe with estimated ages at discovery, 70\% (322) were discovered prior to reaching their peak brightness. As we found in \citet{holoien17a} and \citet{holoien17b}, the ASAS-SN sample remains less affected by host galaxy selection effects than other samples: 24\% (127) of ASAS-SN supernovae occurred in catalogued hosts without previously measured redshifts and an additional 4\% (19) occurred in uncatalogued hosts or were hostless. In contrast, 18\% (74) of non-ASAS-SN supernovae were discovered in known hosts without reported redshifts, and only 2\% (9) were discovered in uncatalogued hosts or were hostless.

Figure~\ref{fig:offmag} shows the $K_S$-band absolute magnitudes of SN host galaxies in our sample compared to the SN offsets from their host nuclei. The Figure shows the ASAS-SN, amateur, and other professional samples as different colors, and also shows the median offsets and magnitudes for each source. To put the magnitude scale in perspective, we also give a corresponding luminosity scale that assumes an $L_\star$ galaxy has $M_{\star,K_S}=-24.2$ \citep{kochanek01}.

Amateur discoveries continue to be biased towards more luminous host galaxies and larger SN offsets \citep{holoien16d,holoien17a,holoien17b}. Supernovae found by other professional surveys continue to exhibit a smaller median angular separation than amateur discoveries (median value of 10\farcs{4} compared to 17\farcs{3}), and now show a smaller median offset in physical separation as well (4.5 kpc vs. 5.7 kpc). This is in contrast to previous years, when the median physical offset was similar between the two groups \citep[e.g.,][]{holoien17b}. ASAS-SN remains less biased against discoveries close to the host nucleus than either group, with the ASAS-SN discoveries showing median offsets of 4\farcs{5} and 2.4 kpc.

The median host galaxy magnitude for ASAS-SN discoveries is $M_{K_S}\simeq -22.7$, compared to $M_{K_S}\simeq -22.9$, and $M_{K_S}\simeq -23.9$ for other professional surveys and amateurs, respectively. This is similar to the trend seen in our previous years' catalogs, where there is a clear distinction in host luminosity between professional surveys (including ASAS-SN) and amateurs \citep{holoien16d, holoien17a,holoien17b}.

Cumulative host galaxy magnitude and SN offset distributions are shown in Figure~\ref{fig:offmag_dist}. The Figure shows clearly that the amateur supernova sample stands out from the ASAS-SN and other professional samples in host galaxy luminosity. It also shows that supernovae discovered by ASAS-SN are more concentrated towards the centers of their hosts, and that those discovered by other professionals fall between the ASAS-SN sample and the amateur sample in offset. A larger fraction of the other professional sample was discovered by surveys that use difference imaging in 2017 compared to previous years, largely due to the ATLAS survey. (See results for previous years in \citealt{holoien16d}, \citealt{holoien17a}, and \citealt{holoien17b}.) While ASAS-SN remains the dominant source of bright supernova discoveries, ATLAS has now surpassed amateur astronomers to become the second largest contributor of bright supernova discoveries. Despite the contribution of ATLAS, ASAS-SN continues to find SNe with smaller median offsets than competing SN searches, and still has a smaller median offset than other professionals (3\farcs{2} vs. 8\farcs{6}) when looking only at 2017 discoveries. This implies that other SN surveys continue to avoid central regions of galaxies in their searches.

In Figure~\ref{fig:histogram} we show the number of bright supernovae discovered in each month from 2012 through 2017 in order to show the impact ASAS-SN has on the discovery of bright supernovae. Various milestones in the history of ASAS-SN, such as the deployment of additional units and software improvements, are also shown in the figure to show the impact of these changes.

%%%%%%%%%%%%%%%%%
% Figure: Redshift Distribution
%%%%%%%%%%%%%%%%%

\begin{figure}
\centering
\subfloat{{\includegraphics[width=0.95\linewidth]{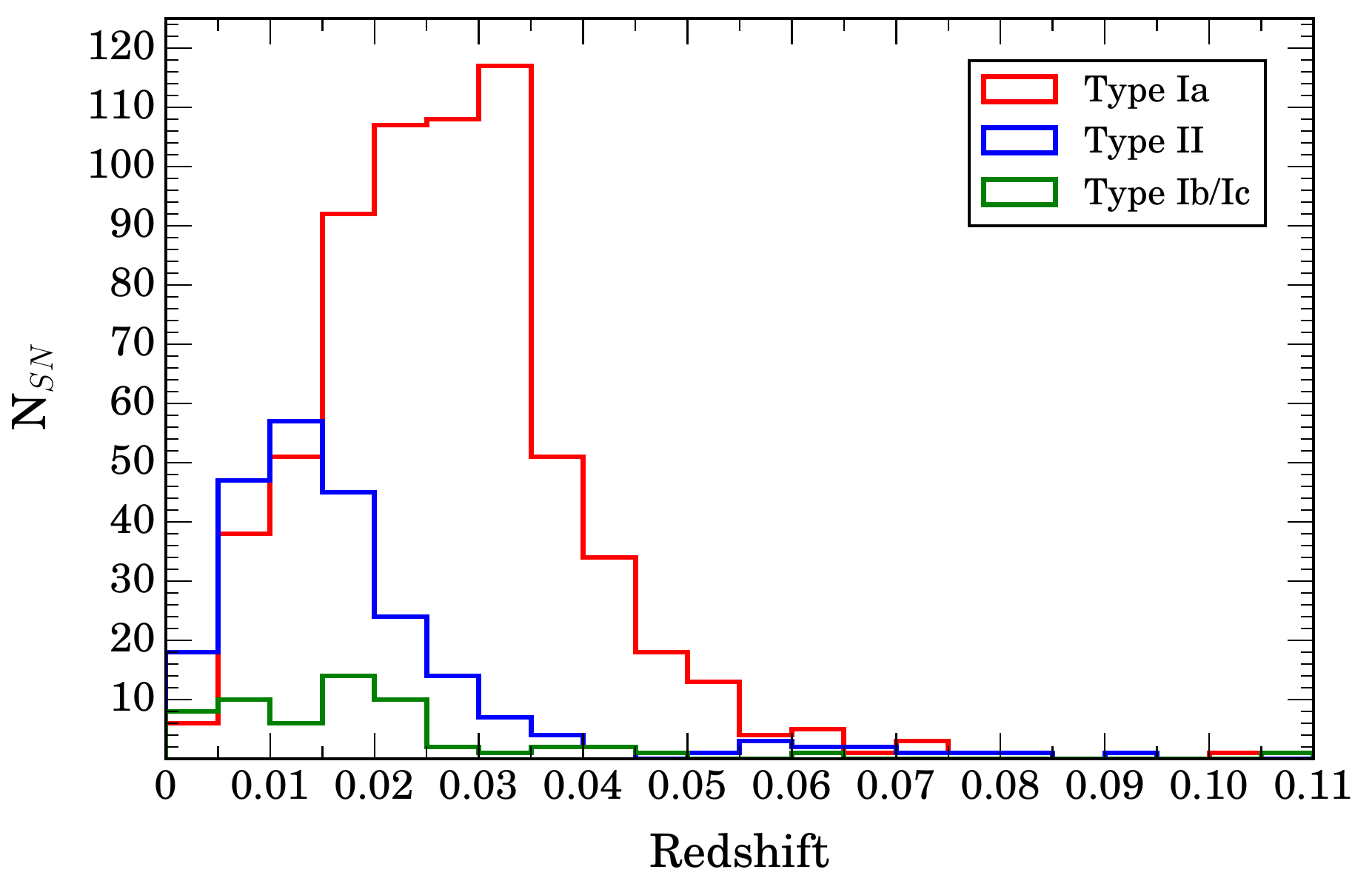}}}
\caption{Histograms of redshifts of the supernovae in our full sample with a bin width of $z=0.005$. We show the distributions for Type Ia (red line), Type II (blue line), and Type Ib/Ic (green line) separately. Type Ia supernovae are largely found at higher redshifts than the core-collapse supernovae, as they are typically more luminous.}
\label{fig:redshift}
\end{figure}

In the months since 2014 May, the average number of bright supernovae found per month is 21 with a scatter of 5 SNe per month. This is an increase compared to an average of 13 with a scatter of 4 SNe per month from 2012 January through 2013 May (the months prior to ASAS-SN beginning operations), and an average of 15 with a scatter of 5 supernovae per month from 2013 June through 2014 May (the months prior to the deployment of our southern unit Cassius), indicating ASAS-SN has increased the rate of bright supernovae found per month since expanding to the southern hemisphere. The addition of the 2017 sample has only improved the trend from \citet{holoien17b}, as the average rate is now $\sim21\pm2$, rather than $\sim20\pm2$ as was seen in the 2016 catalog. While the addition of new surveys like ATLAS has cut into the number of ASAS-SN discoveries somewhat, ASAS-SN still discovered or recovered well over half of all bright supernovae every month in 2017. ASAS-SN continues to find supernovae that would not be found if it did not exist, and this is allowing us to construct a more complete sample of bright supernovae. 

The addition of the new ASAS-SN telescopes did not have a major impact on ASAS-SN discovery numbers in 2017, as all three new units were deployed late in 2017 and were accumulating images with which to build reference images for much of the remainder of the year. We expect an increase in the average number of discoveries in future years because of our increased cadence and coverage, however. For a breakdown of the impact of previous improvements to the ASAS-SN network and pipeline, see Paper III \citep{holoien17b}.

The distribution of redshifts of the supernovae in our total sample, shown in Figure~\ref{fig:redshift}, has not changed significantly from \citet{holoien17b}. The Type Ia distribution peaks between $z=0.03$ and $z=0.035$, the Type II distribution peaks between $z=0.01$ and $z=0.015$, and the Type Ib/Ic distribution peaks between $z=0.015$ and $z=0.02$. As we noted in our previous catalogs, this distribution is not unexpected given that we have a mostly magnitude-limited sample since Type Ia supernovae are more luminous on average than core-collapse supernovae.

\begin{figure}
\centering
\subfloat{{\includegraphics[width=0.95\linewidth]{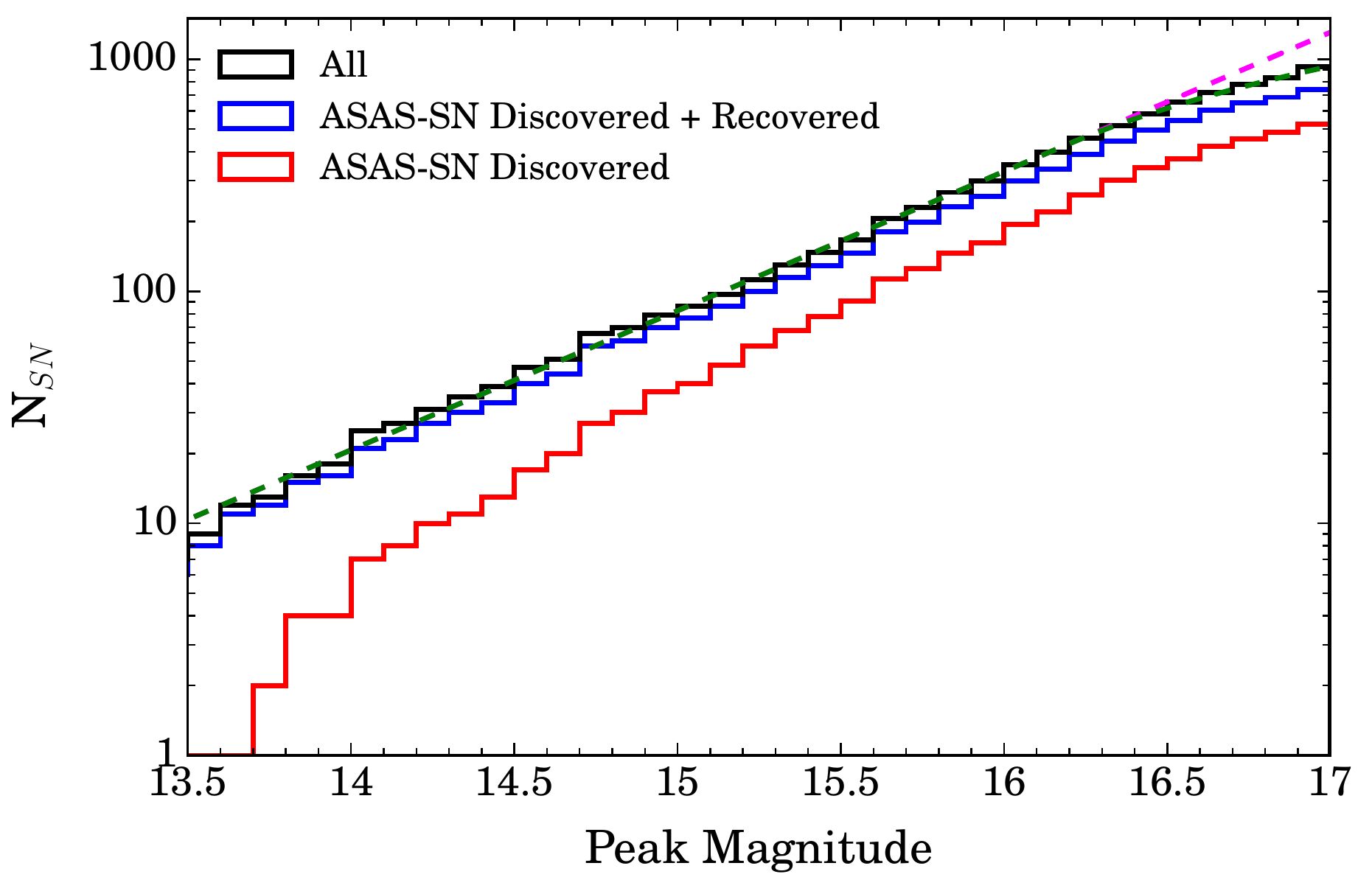}}}
\caption{Cumulative histogram of peak supernovae $V$-band magnitudes. ASAS-SN discoveries are shown with a red line, SNe discovered or recovered by ASAS-SN are shown with a blue line, and edit{the total sample is shown with a black line}. The green dashed line shows a broken power-law fit} that is normalized to the full sample, and has a Euclidean slope below the break magnitude and a variable slope above it. The magenta dashed line shows the Euclidean slope extrapolated to $m=17$, and the sample is approximately 70\% complete for $m_{peak}<17$. Only supernovae with ASAS-SN $V$-band light curves are included in this Figure.
\label{fig:mag_dist}
\end{figure}

Finally, Figure~\ref{fig:mag_dist} shows a cumulative histogram of supernova peak magnitudes for ASAS-SN discoveries, SNe discovered or recovered by ASAS-SN, and all SNe in the total sample. Because the magnitudes from the Bright Supernova website are from different sources and in different filters, for the purposes of this figure we use only supernovae for which we were able to obtain an ASAS-SN $V$-band light curve. This allows us to look at our completeness with a consistent set of peak magnitudes that come from the same telescopes and same filters. This also allows this sample to be more easily applied to supernova rate studies.

The majority of very bright ($m_{peak}\lesssim14.5$; \citealt{holoien17a}) supernovae are discovered by amateurs or the Distance Less Than 40 Mpc (DLT40) survey\footnote{\url{http://dark.physics.ucdavis.edu/dlt40/DLT40}}, both of which typically survey the brightest and nearest galaxies rather than taking an unbiased, all-sky approach. However, ASAS-SN recovers the vast majority of these supernovae, and in 2017 ASAS-SN discovered or recovered all SNe with $m_{peak}<15.3$, a significant improvement from 2016, when we only discovered or recovered everything with $m_{peak}<14.3$ \citep{holoien17b}.

%%%%%%%%%%%%%%%%%
% Figure: Mag Distribution
%%%%%%%%%%%%%%%%%

Figure~\ref{fig:mag_dist} also illustrates the estimated completeness of our sample. We used a broken power-law fit, indicated in the Figure with a green dashed line, to model the distribution of the observable SNe brighter than $m_{peak}=17.01$. This fit assumes a Euclidean slope below the break magnitude and a variable slope above it. We used Markov Chain Monte Carlo methods to derive the parameters of the fit. Similar to what we found in the 2016 catalog, we find that the best-fit break magnitude is $m=16.24\pm0.11$ for the total sample, meaning that the number counts are consistent with a Euclidean slope for magnitudes brighter than 16.24.

The integral completeness of our total sample relative to Euclidean predictions is $0.95\pm0.03$ at $m=16.5$ and $0.73\pm0.03$ at $m=17.0$. The differential completeness relative to Euclidean predictions is $0.71\pm0.07$ at $m=16.5$ and $0.36\pm0.04$ at $m=17.0$. These results are very similar to what we found in \citet{holoien17b} and imply that roughly 70\% of the SNe brighter than $m_{peak}=17$ and roughly 30\% of the $m_{peak}=17$ SNe are being found. We note that the Euclidean approximation we use for this analysis does not account for deviations from Euclidean geometry, K-corrections, or the effects of time dilation on SN rates, so it likely underestimates the true completeness for faint supernovae slightly. We will include these higher order corrections when carrying out an analysis of nearby supernova rates in future work.

%%%%%%%%%%%%%%%%%
% Section: Conclusions
%%%%%%%%%%%%%%%%%

\section{Conclusions}
\label{sec:disc}

This manuscript presents a catalog of bright, spectroscopically confirmed SNe and their hosts, totaling 308 new supernovae discovered in 2017. Our full combined bright SN sample now includes 949 SNe, with 528 of these discovered by ASAS-SN and an additional 216 independently recovered by ASAS-SN after discovery. The ASAS-SN sample continues to closely resemble an ideal magnitude-limited sample defined by \citet{li11}, while the combined sample has a similar distribution but with a larger proportion of core-collapse SNe relative to Type Ia SNe than expected.

ASAS-SN remains the only professional survey to provide a rapid-cadence, all-sky survey of bright transients, and with the expansion of our telescope network in 2017, it will have an even greater impact on the discovery of bright SNe going forward. The ATLAS survey is now the primary competitor to ASAS-SN for new bright SN discoveries, though amateur astronomers still discover a significant fraction of the brightest SNe and discover the third most SNe in our 2017 sample overall. Despite the advent of recent professional surveys like ATLAS, ASAS-SN continues to find SNe closer to galactic nuclei than its competitors (Figure~\ref{fig:offmag}), and finds SNe that would not be found otherwise (Figure~\ref{fig:histogram}). As in 2015 and 2016, ASAS-SN recovered the majority of non-ASAS-SN discoveries in 2017, and it recovered or discovered all supernovae with $m_{peak}\leq15.3$ in 2017.

Similar to what we found in \citet{holoien17b}, our total sample is roughly complete to a peak magnitude of $m_{peak}=16.2$, and is roughly 70\% complete for $m_{peak}\leq17.0$. This analysis served as the precursor to our first supernova rates paper, where we found that the specific Type Ia supernova rate rises as host galaxy mass decreases \citep{brown18}. Manuscripts with further rate calculations with respect to other host properties (e.g., star formation rate and metallicity) and an overall nearby supernova rate paper are in preparation, and will have a significant impact on a number of fields of astronomy, including the nearby core-collapse rate \citep[e.g.,][]{horiuchi11,horiuchi13} and multi-messenger studies ranging from gravitational waves \citep[e.g.,][]{ando13,nakamura16}, to MeV gamma rays from Type Ia supernovae \citep[e.g.,][]{horiuchi10,diehl14,churazov15} to GeV--TeV gamma rays and neutrinos from rare types of core-collapse supernovae \citep[e.g.,][]{ando05,murase11,abbasi12}. Joint measurements such as these will greatly increase the scientific reach of ASAS-SN.

This is the fourth of a yearly series of catalogs of bright SNe and their hosts provided by the ASAS-SN team. These catalogs are intended to provide useful data repositories for bright SN and host galaxy properties that can be used for new and interesting population studies. As ASAS-SN continues to dominate the discovery of the best and brightest transients in the sky, this is just one way in which we can leverage our unbiased sample to improve supernova research.

\section*{Acknowledgments}

The authors thank Las Cumbres Observatory and its staff for their continued support of ASAS-SN.  

ASAS-SN is supported by the Gordon and Betty Moore Foundation through grant GBMF5490 to the Ohio State University and NSF grant AST-1814440. Development of ASAS-SN has been supported by NSF grant AST-0908816, the Center for Cosmology and AstroParticle Physics at the Ohio State University, the Mt. Cuba Astronomical Foundation, the Chinese Academy of Sciences South America Center for Astronomy (CASSACA), and by George Skestos.

This material is based upon work supported by the National Science Foundation Graduate Research Fellowship Program Under Grant No. DGE-1343012. JSB, KZS, and CSK are supported by NSF grant AST-181440. KZS and CSK are also supported by NSF grant AST-1515876. Support for JLP is provided in part by FONDECYT through the grant 1151445 and by the Ministry of Economy, Development, and Tourism's Millennium Science Initiative through grant IC120009, awarded to The Millennium Institute of Astrophysics, MAS. SD, SB, and PC acknowledge Project 11573003 supported by NSFC. JFB is supported by NSF grant PHY-1714479. MDS is supported by generous grants from the Villum foundation (grant 13261) and the Independent Research Fund Denmark. TAT is supported in part by Scialog Scholar grant 24215 from the Research Corporation. PRW acknowledges support from the US Department of Energy as part of the Laboratory Directed Research and Development program at LANL. 

This research uses data obtained through the Telescope Access Program (TAP), which has been funded by ``the Strategic Priority Research Program-The Emergence of Cosmological Structures'' of the Chinese Academy of Sciences (Grant No.11 XDB09000000) and the Special Fund for Astronomy from the Ministry of Finance.

This research has made use of the XRT Data Analysis Software (XRTDAS) developed under the responsibility of the ASI Science Data Center (ASDC), Italy. At Penn State the NASA {\swift} program is support through contract NAS5-00136.

This research was made possible through the use of the AAVSO Photometric All-Sky Survey (APASS), funded by the Robert Martin Ayers Sciences Fund.

This research has made use of data provided by Astrometry.net \citep{barron08,lang10}.

This paper uses data products produced by the OIR Telescope Data Center, supported by the Smithsonian Astrophysical Observatory.

Observations made with the NASA Galaxy Evolution Explorer (GALEX) were used in the analyses presented in this manuscript. Some of the data presented in this paper were obtained from the Mikulski Archive for Space Telescopes (MAST). STScI is operated by the Association of Universities for Research in Astronomy, Inc., under NASA contract NAS5-26555. Support for MAST for non-HST data is provided by the NASA Office of Space Science via grant NNX13AC07G and by other grants and contracts.

Funding for SDSS-III has been provided by the Alfred P. Sloan Foundation, the Participating Institutions, the National Science Foundation, and the U.S. Department of Energy Office of Science. The SDSS-III web site is http://www.sdss3.org/.

This publication makes use of data products from the Two Micron All Sky Survey, which is a joint project of the University of Massachusetts and the Infrared Processing and Analysis Center/California Institute of Technology, funded by NASA and the National Science Foundation.

This publication makes use of data products from the Wide-field Infrared Survey Explorer, which is a joint project of the University of California, Los Angeles, and the Jet Propulsion Laboratory/California Institute of Technology, funded by NASA.

This research is based in part on observations obtained at the Southern Astrophysical Research (SOAR) telescope, which is a joint project of the Minist\'{e}rio da Ci\^{e}ncia, Tecnologia, e Inova\c{c}\~{a}o (MCTI) da Rep\'{u}blica Federativa do Brasil, the U.S. National Optical Astronomy Observatory (NOAO), the University of North Carolina at Chapel Hill (UNC), and Michigan State University (MSU). 

The Liverpool Telescope is operated on the island of La Palma by Liverpool John Moores University in the Spanish Observatorio del Roque de los Muchachos of the Instituto de Astrofisica de Canarias with financial support from the UK Science and Technology Facilities Council.

This research has made use of the NASA/IPAC Extragalactic Database (NED), which is operated by the Jet Propulsion Laboratory, California Institute of Technology, under contract with NASA.

\bibliographystyle{mnras2}
\bibliography{bibliography_catalogs}

\newpage

%%%%%%%%%%%%%%%%%
% Table: ASAS-SN Supernovae
%%%%%%%%%%%%%%%%%
\begin{landscape}
\begin{table}
\begin{minipage}{\textwidth}
\centering
\fontsize{6}{7.2}\selectfont
\caption{ASAS-SN Supernovae}
\label{table:asassn_sne}
\begin{tabular}{@{}l@{\hspace{0.15cm}}l@{\hspace{0.15cm}}c@{\hspace{0.15cm}}c@{\hspace{0.15cm}}c@{\hspace{0.15cm}}l@{\hspace{0.15cm}}c@{\hspace{0.15cm}}c@{\hspace{0.15cm}}c@{\hspace{0.15cm}}c@{\hspace{0.15cm}}c@{\hspace{0.15cm}}c@{\hspace{0.15cm}}l@{\hspace{0.15cm}}l@{\hspace{0.15cm}}l@{\hspace{-0.05cm}}} 
\hline
\vspace{-0.14cm}
 & & & & & & & & & & & & & & \\
 & IAU & Discovery & & & & & & & Offset & & Age & & & \\
SN Name & Name & Date & RA$^a$ & Dec.$^a$ & Redshift & $m_{disc}^b$ & $V_{peak}^c$ & $g_{peak}^c$ & (arcsec)$^d$ & Type & at Disc.$^e$ & Host Name$^f$ & Discovery ATel & Classification ATel \\
\vspace{-0.23cm} \\
\hline
\vspace{-0.17cm}
 & & & & & & & & & & & & & &\\
ASASSN-17ac  &  2017ad  &  2017-01-04.36  &  14:34:26.01  &  $-$38:28:09.70  &  0.03332  &  16.6  &  16.3  &  ---  &  6.09  &  Ia  &  $-6$  &  2MASX J14342552  & \citet{asassn17ac_atel} & \citet{asassn17ac_spec_atel} \\ 
ASASSN-17ad  &  2017ah  &  2017-01-04.55  &  11:10:01.95  &  $+$63:38:34.16  &  0.03286  &  17.3  &  15.9  &  ---  &  3.82  &  Ia  &  $-11$  &  CGCG 314-011  & \citet{asassn17ad_atel} & TNS \\
ASASSN-17ae  &  2017ai  &  2017-01-04.66  &  16:17:02.62  &  $+$10:41:36.17  &  0.05027  &  17.5  &  17.6  &  ---  &  11.50  &  Ia  &  $-3$  &  2MASX J16170338  & \citet{asassn17ae_atel} & TNS \\ 
ASASSN-17af  &  2017bc  &  2017-01-05.51  &  12:19:50.90  &  $-$06:51:20.45  &  0.02687  &  17.0  &  16.4  &  ---  &  4.47  &  Ia  &  $-4$  &  MCG -01-32-001  & \citet{asassn17af_atel} & \citet{asassn17ac_spec_atel} \\ 
ASASSN-17ai  &  2017hl  &  2017-01-09.63  &  12:07:18.83  &  $+$16:50:26.02  &  0.02307  &  17.3  &  16.7  &  ---  &  4.74  &  Ib  &  $-7$  &  KUG 1204+171  & \citet{asassn17ai_atel} & \citet{asassn17ai_spec_atel} \\ 
ASASSN-17aj  &  2017hm  &  2017-01-09.62  &  11:33:10.50  &  $-$10:13:18.37  &  0.02128  &  16.9  &  15.8  &  ---  &  25.50  &  Ia  &  $-13$  &  MCG -02-30-003  & \citet{asassn17ai_atel} & \citet{asassn17ai_spec_atel} \\ 
ASASSN-17am  &  2017hq  &  2017-01-10.66  &  13:49:23.81  &  $+$08:30:27.62  &  0.03798  &  17.6  &  16.5  &  ---  &  2.21  &  Ia  &  18  &  CGCG 073-079  & \citet{asassn17am_atel} & TNS \\ 
ASASSN-17ap  &  2017je  &  2017-01-03.11  &  00:37:37.59  &  $-$34:29:49.24  &  0.04500  &  17.4  &  16.8  &  ---  &  8.94  &  Ia  &  $-10$  &  GALEXASC J003737 & \citet{asassn17ap_atel} & \citet{asassn17ap_spec_atel} \\ 
ASASSN-17at  &  2017ln  &  2017-01-19.54  &  11:38:33.66  &  $+$25:23:50.17  &  0.02536  &  16.7  &  16.4  &  ---  &  3.08  &  Ia  &  $-9$  &  2MASX J11383367  & \citet{asassn17at_atel} & \citet{asassn17at_spec_atel} \\ 
ASASSN-17bb  &  2017ng  &  2017-01-23.65  &  15:20:40.75  &  $+$04:39:34.42  &  0.03700  &  16.9  &  16.7  &  ---  &  1.86  &  Ia  &  $-3$  &  2MASX J15204087  & \citet{asassn17bb_atel} & \citet{asassn17ap_spec_atel} \\ 
ASASSN-17bc  &  2017nh  &  2017-01-23.36  &  07:10:13.52  &  $+$27:12:09.97  &  0.06100  &  17.6  &  16.8  &  ---  &  6.18  &  Ia  &  1  &  2MASX J07101346  & \citet{asassn17bc_atel} & \citet{asassn17bc_spec_atel} \\ 
ASASSN-17bd  &  2017nk  &  2017-01-23.61  &  15:59:18.43  &  $+$13:36:50.89  &  0.03455  &  17.3  &  17.1  &  ---  &  3.00  &  Ia  &  3  &  2MASX J15591858  & \citet{asassn17bd_atel} & \citet{asassn17ap_spec_atel} \\ 
ASASSN-17be  &  2017pa  &  2017-01-17.07  &  02:03:10.53  &  $-$61:41:10.61  &  0.04000  &  17.5  &  17.0  &  ---  &  0.51  &  Ia  &  $-1$  &  2MASX J02031063  & \citet{asassn17bd_atel} & \citet{asassn17bd_atel} \\ 
ASASSN-17bh  &  2017po  &  2017-01-26.61  &  16:03:51.70  &  $+$39:59:24.17  &  0.03186  &  16.6  &  15.8  &  ---  &  14.52  &  Ia  &  $-3$  &  CGCG 223-033  & \citet{asassn17ap_spec_atel} & \citet{asassn17bh_spec_atel} \\ 
ASASSN-17bn  &  2017vu  &  2017-01-21.43  &  08:59:23.92  &  $-$09:52:29.32  &  0.04451  &  17.6  &  17.0  &  ---  &  0.61  &  Ia  &  $-2$  &  2MASX J08592386  & \citet{asassn17bh_spec_atel} & \citet{asassn17bh_spec_atel} \\ 
ASASSN-17bo  &  2017wb  &  2017-01-28.59  &  11:01:19.53  &  $+$70:39:54.76  &  0.03000  &  16.9  &  16.3  &  ---  &  1.91  &  Ia  &  $-9$  &  2MASX J11011991  & \citet{asassn17bo_atel} & \citet{asassn17bo_atel} \\ 
ASASSN-17bp  &  2017wi  &  2017-01-29.05  &  02:02:08.63  &  $-$17:59:56.36  &  0.05100  &  17.2  &  17.0  &  ---  &  3.45  &  Ia  &  7  &  GALEXASC J020208  & \citet{asassn17bo_atel} & \citet{asassn17bp_spec_atel} \\ 
ASASSN-17bq  &  2017xx  &  2017-01-27.51  &  07:25:38.21  &  $+$59:00:09.63  &  0.04000  &  17.6  &  16.4  &  ---  &  1.51  &  Ia  &  19  &  GALEXASC J072538  & \citet{asassn17bo_atel} & \citet{asassn17bo_atel} \\ 
ASASSN-17br  &  2017xy  &  2017-01-29.63  &  15:52:00.31  &  $+$66:18:55.27  &  0.02600  &  17.1  &  18.1  &  ---  &  3.71  &  IIP  &  10  &  GALEXASCJ155200  & \citet{asassn17bo_atel} & \citet{asassn17bh_spec_atel} \\ 
ASASSN-17bs  &  2017yh  &  2017-01-30.66  &  17:52:06.13  &  $+$21:33:57.82  &  0.02040  &  16.5  &  15.9  &  ---  &  13.02  &  Ia  &  $-8$  &  IC 1269  & \citet{asassn17bh_spec_atel} & \citet{asassn17bp_spec_atel} \\
\vspace{-0.22cm}
 & & & & & & & & & & & & & &\\
\hline
\end{tabular}
\smallskip
\\
\raggedright
\noindent This table is available in its entirety in a machine-readable form in the online journal. A portion is shown here for guidance regarding its form and content.\\
$^a$ Right ascension and declination are given in the J2000 epoch. \\
$^b$ Discovery magnitudes are $V$- or $g$-band magnitudes from ASAS-SN, depending on the camera used for discovery. \\
$^c$ Peak $V$- and $g$-band magnitudes are measured from ASAS-SN data. \\
$^d$ Offset indicates the offset of the supernova in arcseconds from the coordinates of the host nucleus, taken from NED. \\
$^e$ Discovery ages are given in days relative to peak. All ages are approximate and are only listed if a clear age was given in the classification telegram. \\
$^e$ ``2MASSX'' and ``GALEXASC'' host names have been abbreviated due to space constrtaints. \\
\vspace{-0.5cm}
\end{minipage}
\end{table}

%%%%%%%%%%%%%%%%%
% Table: Other Supernovae
%%%%%%%%%%%%%%%%%

\begin{table}
\begin{minipage}{\textwidth}
\bigskip\bigskip
\centering
\fontsize{6}{7.2}\selectfont
\caption{Non-ASAS-SN Supernovae}
\label{table:other_sne}
\begin{tabular}{@{}l@{\hspace{0.15cm}}l@{\hspace{0.15cm}}c@{\hspace{0.15cm}}c@{\hspace{0.15cm}}c@{\hspace{0.15cm}}l@{\hspace{0.15cm}}c@{\hspace{0.15cm}}c@{\hspace{0.15cm}}c@{\hspace{0.15cm}}c@{\hspace{0.15cm}}c@{\hspace{0.15cm}}l@{\hspace{0.15cm}}c@{\hspace{0.15cm}}c} 
\hline
\vspace{-0.14cm}
 & & & & & & & & & & & & & \\
 & IAU & Discovery &  & & & & & & Offset & & & & \\
 SN Name & Name & Date & RA$^a$ & Dec.$^a$ & Redshift & $m_{peak}^b$ & $V_{peak}^c$ & $g_{peak}^c$ & (arcsec)$^d$ & Type & Host Name & Discovered By$^e$ & Recovered?$^f$ \\
\vspace{-0.23cm} \\
\hline
\vspace{-0.17cm}
 & & & & & & & & & & & \\
ATLAS17abh & 2017ae & 2017-01-04.29 & 02:05:50.62 & $+$18:22:30.23 & 0.022000 & 16.4 & 16.2 & --- & 4.74 & Ia & GALEXASC J020550 & ATLAS & Yes \\ 
2017hr & 2017hr & 2017-01-06.69 & 12:06:27.39 & $+$28:08:19.70 & 0.029300 & 16.6 & 17.0 & --- & 1.38 & Ia & SDSS J120627.46 & Amateurs & Yes \\ 
PS17hj & 2017jd & 2017-01-09.21 & 23:34:36.47 & $-$04:32:04.32 & 0.007368 & 14.6 & --- & --- & 1.26 & Ia & IC 5334 & Pan-STARRS & No \\ 
2017hn & 2017hn & 2017-01-09.41 & 13:07:39.46 & $+$06:20:14.60 & 0.023853 & 16.1 & 15.8 & --- & 4.66 & Ia & UGC 08204 & Amateurs & Yes \\ 
ATLAS17ajn & 2017lv & 2017-01-14.63 & 11:44:26.54 & $-$28:27:27.22 & 0.028717 & 17.0 & 16.7 & --- & 18.48 & Ia & ESO 440-G001 & ATLAS & Yes \\ 
MASTER OT J081506.13+381123.3 & --- & 2017-01-16.02 & 08:15:6.13 & $+$38:11:23.30 & 0.054000 & 16.9 & 16.6 & --- & 11.35 & Ia & 2MASX J08150520 & MASTER & No \\ 
ATLAS17air & 2017jl & 2017-01-16.22 & 00:57:31.90 & $+$30:11:06.83 & 0.016331 & 14.6 & 15.3 & --- & 5.88 & Ia & 2MASX J00573150 & ATLAS & Yes \\ 
2017mf & 2017mf & 2017-01-21.09 & 14:16:31.0 & $+$39:35:12.02 & 0.025678 & 16.0 & 15.9 & --- & 12.6 & Ia & NGC 5541 & Amateurs & Yes \\ 
PTSS-17dfc & 2017ms & 2017-01-21.71 & 10:26:42.37 & $+$36:40:50.62 & 0.024639 & 15.6 & 15.7 & --- & 5.22 & Ia & SDSS J102641.99 & PTSS & Yes \\ 
ATLAS17akw & 2017nt & 2017-01-23.21 & 23:53:31.13 & $+$03:44:08.18 & 0.038800 & 16.8 & --- & --- & 34.38 & Ia-91T & SSTSL2 J235328.89 & ATLAS & No \\ 
ATLAS17alb & 2017ns & 2017-01-23.31 & 02:49:10.36 & $+$14:36:02.48 & 0.027900 & 16.7 & --- & --- & 2.88 & Ia & 2MASX J02491020 & ATLAS & No \\ 
ATLAS17amz & 2017pn & 2017-01-26.31 & 04:46:24.59 & $-$11:59:18.25 & 0.014000 & 16.5 & 15.8 & --- & 0 & IIP & Uncatalogued & ATLAS & No \\ 
ATLAS17auc & 2017zd & 2017-01-26.64 & 13:32:42.09 & $-$21:48:04.35 & 0.02947 & 15.9 & 15.9 & --- & 1.38 & Ia & 2MASX J13324217 & ATLAS & Yes \\ 
ATLAS17axb & 2017adj & 2017-01-30.65 & 13:43:23.25 & $-$19:56:37.13 & 0.030000 & 16.9 & 16.1 & --- & 3.84 & Ia & GALEXASC J134322 & ATLAS & Yes \\ 
Gaia17aiq & 2017ati & 2017-02-06.43 & 09:49:56.70 & $+$67:10:59.56 & 0.01305 & 16.0 & 16.2 & --- & 38.6 & IIb & KUG 0945+674 & Gaia & Yes \\ 
DLT17h & 2017ahn & 2017-02-08.36 & 10:37:17.45 & $-$41:37:05.27 & 0.009255 & 15.8 & 15.3 & --- & 40.32 & II & NGC 3318 & DLT40 & Yes \\ 
MASTER OT J083256.92-035128.1 & 2017cgr & 2017-02-11.93 & 08:32:56.92 & $-$03:51:28.10 & 0.030584 & 16.5 & 16.5 & --- & 5.82 & Ia & 2MASX J08325728 & MASTER & Yes \\ 
PS17bbn & 2017avj & 2017-02-14.58 & 13:05:3.01 & $+$53:39:33.21 & 0.029037 & 17.0 & 16.7 & --- & 19.44 & Ia-91bg & CGCG 270-047 & Pan-STARRS & Yes \\ 
iPTF17aub & 2017aub & 2017-02-15.3 & 06:40:24.70 & $+$64:33:02.75 & 0.016000 & 17.0 & 16.9 & --- & 10.02 & II & CGCG 308-036 & PTF & No \\ 
ATLAS17bam & 2017avl & 2017-02-16.32 & 05:20:47.08 & $+$03:15:24.55 & 0.027426 & 16.4 & 16.4 & --- & 23.46 & Ia & CGCG 421-034 & ATLAS & Yes \\  
\vspace{-0.22cm}
 & & & & & & & & & & & \\
\hline
\end{tabular}
\smallskip
\\
\raggedright
\noindent This table is available in its entirety in a machine-readable form in the online journal. A portion is shown here for guidance regarding its form and content.\\
$^b$ Right ascension and declination are given in the J2000 epoch. \\
$^c$ Magnitudes are taken from D. W. Bishop's Bright Supernova website, as described in the text, and may be from different filters. \\
$^d$ All $V-$ and $g-$band peak magnitudes are measured from ASAS-SN data for cases where the supernova was detected. \\
$^e$ Offset indicates the offset of the supernovae in arcseconds from the coordinates of the host nucleus, taken from NED. \\
$^f$ ``Amateurs'' indicates discovery by any number of non-professional astronomers, as described in the text. \\
$^g$ Indicates whether the supernova was independently recovered in ASAS-SN data or not.
\end{minipage}
\vspace{-0.5cm}
\end{table}

\end{landscape}
\pagebreak
\begin{landscape}

%%%%%%%%%%%%%%%%%
% Table: ASASSN Hosts
%%%%%%%%%%%%%%%%%

\begin{table}
\begin{minipage}{\textwidth}
\centering
\fontsize{6}{7.2}\selectfont
\caption{ASAS-SN Supernova Host Galaxies}
\label{table:asassn_hosts}
\begin{tabular}{@{}l@{\hspace{0.15cm}}l@{\hspace{0.15cm}}c@{\hspace{0.15cm}}c@{\hspace{0.15cm}}c@{\hspace{0.15cm}}c@{\hspace{0.15cm}}c@{\hspace{0.15cm}}c@{\hspace{0.15cm}}c@{\hspace{0.15cm}}c@{\hspace{0.15cm}}c@{\hspace{0.15cm}}c@{\hspace{0.15cm}}c@{\hspace{0.15cm}}c@{\hspace{0.15cm}}c@{\hspace{0.15cm}}c@{\hspace{0.15cm}}c} 
\hline
\vspace{-0.14cm}
 & & & & & & & & & & & & & & \\
 & & SN & SN & SN Offset & & & & & & & & & & \\
Galaxy Name & Redshift & Name & Type & (arcsec) & $A_V$$^a$ & $m_{NUV}$$^b$ & $m_u$$^c$ & $m_g$$^c$ & $m_r$$^c$ & $m_i$$^c$ & $m_z$$^c$ & $m_J$$^d$ & $m_H$$^d$ & $m_{K_S}$$^{d,e}$ & $m_{W1}$ & $m_{W2}$\\ 
\vspace{-0.23cm} \\
\hline
\vspace{-0.17cm}
 & & & & & & & & & & & & & & \\
2MASX J14342552-3828081 & 0.03332 & ASASSN-17ac & Ia & 6.09 & 0.280 & --- & --- & --- & --- & --- & --- & 13.56 0.07 & 12.74 0.07 & 12.41 0.11 & 12.83 0.02 & 12.87 0.03 \\ 
CGCG 314-011 & 0.03286 & ASASSN-17ad & Ia & 3.82 & 0.032 & 19.70 0.12 & 16.05 0.01 & 14.29 0.00 & 13.51 0.00 & 13.12 0.00 & 12.86 0.00 & 11.89 0.02 & 11.18 0.04 & 10.87 0.05 & 11.31 0.02 & 11.36 0.02 \\ 
2MASX J16170338+1041359 & 0.05027 & ASASSN-17ae & Ia & 11.50 & 0.166 & 19.02 0.07 & 17.81 0.04 & 16.41 0.00 & 15.76 0.00 & 15.40 0.00 & 15.17 0.02 & 14.46 0.12 & 13.70 0.12 & 13.25 0.18 & 14.10 0.07 & 13.98 0.04 \\ 
MCG -01-32-001 & 0.02687 & ASASSN-17af & Ia & 4.47 & 0.102 & 16.92 0.02 & --- & --- & --- & --- & --- & 11.88 0.03 & 11.13 0.04 & 10.85 0.06 & 11.38 0.02 & 11.37 0.02 \\ 
KUG 1204+171 & 0.02307 & ASASSN-17ai & Ib & 4.74 & 0.129 & 17.50 0.04 & 16.60 0.01 & 15.55 0.00 & 15.09 0.00 & 14.88 0.00 & 14.66 0.01 & 13.81 0.05 & 13.09 0.07 & 12.97 0.09 & 13.09 0.03 & 12.95 0.03 \\ 
MCG -02-30-003 & 0.02128 & ASASSN-17aj & Ia & 25.50 & 0.114 & 17.80 0.03 & --- & --- & --- & --- & --- & 12.37 0.04 & 11.67 0.05 & 11.27 0.08 & 11.80 0.02 & 11.77 0.02 \\ 
CGCG 073-079 & 0.03798 & ASASSN-17am & Ia & 2.21 & 0.077 & 17.17 0.04 & 16.22 0.01 & 14.53 0.00 & 13.71 0.00 & 13.33 0.00 & 12.98 0.00 & 12.18 0.04 & 11.44 0.05 & 11.18 0.08 & 12.05 0.02 & 12.13 0.03 \\ 
GALEXASC J003737.20-342957.7 & 0.04500 & ASASSN-17ap & Ia & 8.94 & 0.037 & 19.43 0.05 & --- & --- & --- & --- & --- & $>$17.0 & $>$16.4 & 15.44 0.06* & 15.95 0.05 & 15.94 0.14 \\ 
2MASX J11383367+2523532 & 0.02536 & ASASSN-17at & Ia & 3.08 & 0.075 & --- & 16.83 0.01 & 15.38 0.00 & 14.66 0.00 & 14.27 0.00 & 14.01 0.00 & 12.85 0.03 & 12.18 0.05 & 11.80 0.06 & 12.15 0.02 & 11.57 0.02 \\ 
2MASX J15204087+0439331 & 0.03000 & ASASSN-17bb & Ia & 1.86 & 0.121 & 19.07 0.08 & 18.49 0.06 & 17.17 0.01 & 16.61 0.01 & 16.27 0.01 & 16.06 0.01 & 14.90 0.08 & 14.30 0.10 & 13.97 0.15 & 13.94 0.03 & 13.84 0.04 \\ 
2MASX J07101346+2712041 & 0.06100 & ASASSN-17bc & Ia & 6.18 & 0.169 & --- & --- & --- & --- & --- & --- & 13.26 0.03 & 12.62 0.05 & 12.28 0.06 & 12.71 0.04 & 12.72 0.05 \\ 
2MASX J15591858+1336487 & 0.03455 & ASASSN-17bd & Ia & 3.00 & 0.129 & 17.42 0.02 & 16.46 0.01 & 15.65 0.00 & 15.24 0.00 & 15.00 0.00 & 14.81 0.01 & 13.61 0.06 & 12.93 0.07 & 12.46 0.10 & 12.55 0.04 & 12.27 0.03 \\ 
2MASX J02031063-6141105 & 0.04000 & ASASSN-17be & Ia & 0.51 & 0.096 & 21.32 0.28 & --- & --- & --- & --- & --- & 12.98 0.04 & 12.26 0.05 & 11.90 0.08 & 12.12 0.02 & 12.17 0.02 \\ 
CGCG 223-033 & 0.03186 & ASASSN-17bh & Ia & 14.52 & 0.038 & 18.99 0.06 & --- & --- & --- & --- & --- & $>$17.0 & $>$16.4 & $>$15.6 & --- & --- \\ 
2MASX J08592386-0952291 & 0.04451 & ASASSN-17bn & Ia & 0.61 & 0.101 & 21.08 0.21 & --- & --- & --- & --- & --- & 12.70 0.03 & 12.03 0.03 & 11.69 0.05 & 11.84 0.02 & 11.80 0.02 \\ 
2MASX J11011991+7039548 & 0.03000 & ASASSN-17bo & Ia & 1.91 & 0.068 & 18.14 0.05 & --- & --- & --- & --- & --- & 14.81 0.08 & 14.22 0.11 & 13.87 0.13 & 13.80 0.03 & 13.69 0.03 \\ 
GALEXASC J020208.73-175958.3 & 0.05100 & ASASSN-17bp & Ia & 3.45 & 0.075 & 18.68 0.04 & 17.86 0.02 & 16.92 0.01 & 16.63 0.01 & 16.41 0.01 & 16.30 0.02 & $>$17.0 & $>$16.4 & 14.27 0.05* & 14.78 0.03 & 14.53 0.05 \\ 
GALEXASC J072538.14+590010.5 & 0.04000 & ASASSN-17bq & Ia & 1.51 & 0.136 & 21.59 0.36 & --- & --- & --- & --- & --- & 14.24 0.05 & 13.55 0.07 & 13.20 0.08 & 13.83 0.03 & 13.81 0.04 \\ 
GALEXASCJ155200.16+661851.6 & 0.02600 & ASASSN-17br & IIP & 3.71 & 0.075 & --- & --- & --- & --- & --- & --- & $>$17.0 & $>$16.4 & 15.07 0.05* & 15.58 0.03 & 15.73 0.07 \\ 
IC 1269 & 0.02040 & ASASSN-17bs & Ia & 13.02 & 0.243 & 16.11 0.01 & --- & --- & --- & --- & --- & 11.83 0.03 & 11.19 0.05 & 10.89 0.06 & 11.95 0.02 & 11.82 0.02 \\
\vspace{-0.22cm}
 & & & & & & & & & & & & & & \\
\hline
\end{tabular}
\smallskip
\\
\raggedright
\noindent This table is available in its entirety in a machine-readable form in the online journal. A portion is shown here for guidance regarding its form and content. Uncertainty is given for all magnitudes, and in some cases is equal to zero.\\
$^a$ Galactic extinction taken from \citet{schlafly11}. \\
$^b$ No magnitude is listed for those galaxies not detected in GALEX survey data. \\
$^c$ No magnitude is listed for those galaxies not detected in SDSS data or those located outside of the SDSS footprint. \\
$^d$ For those galaxies not detected in 2MASS data, we assume an upper limit of the faintest galaxy detected in each band from our sample. \\
$^e$ $K_S$-band magnitudes marked with a ``*'' indicate those estimated from the WISE $W1$-band data, as described in the text. \\
\end{minipage}
\vspace{-0.5cm}
\end{table}

%%%%%%%%%%%%%%%%%
% Table: Non-ASASSN Hosts
%%%%%%%%%%%%%%%%%

\begin{table}
\begin{minipage}{\textwidth}
\bigskip\bigskip
\centering
\fontsize{6}{7.2}\selectfont
\caption{Non-ASAS-SN Supernova Host Galaxies}
\label{table:other_hosts}
\begin{tabular}{@{}l@{\hspace{0.15cm}}l@{\hspace{0.15cm}}c@{\hspace{0.15cm}}c@{\hspace{0.05cm}}c@{\hspace{0.15cm}}c@{\hspace{0.15cm}}c@{\hspace{0.15cm}}c@{\hspace{0.15cm}}c@{\hspace{0.15cm}}c@{\hspace{0.15cm}}c@{\hspace{0.15cm}}c@{\hspace{0.15cm}}c@{\hspace{0.15cm}}c@{\hspace{0.15cm}}c@{\hspace{0.15cm}}c@{\hspace{0.15cm}}c} 
\hline
\vspace{-0.14cm}
 & & & & & & & & & & & & & & \\
 & & SN & SN & SN Offset & & & & & & & & & & \\
Galaxy Name & Redshift & Name & Type & (arcsec) & $A_V^a$ & $m_{NUV}^b$ & $m_u^c$ & $m_g^c$ & $m_r^c$ & $m_i^c$ & $m_z^c$ & $m_J^d$ & $m_H^d$ & $m_{K_S}^{d,e}$ & $m_{W1}$ & $m_{W2}$ \\ 
\vspace{-0.23cm} \\
\hline
\vspace{-0.17cm}
 & & & & & & & & & & & & & & \\
GALEXASC J020550.53 & 0.022000 & ATLAS17abh & Ia & 4.74 & 0.203 & 21.26 0.27 & 20.25 0.22 & 18.36 0.02 & 17.71 0.02 & 17.35 0.01 & 17.13 0.04 & $>$17.0 & $>$16.4 & 15.17 0.06* & 15.68 0.04 & 15.79 0.13 \\ 
SDSS J120627.46+280820.6 & 0.029300 & 2017hr & Ia & 1.38 & 0.061 & --- & 21.49 0.13 & 20.48 0.03 & 19.58 0.02 & 19.14 0.02 & 18.81 0.05 & $>$17.0 & $>$16.4 & 15.69 0.07* & 16.20 0.06 & 16.04 0.17 \\ 
IC 5334 & 0.007368 & PS17hj & Ia & 1.26 & 0.114 & 16.96 0.04 & 15.22 0.01 & 13.47 0.00 & 12.70 0.00 & 12.28 0.00 & 11.96 0.00 & 11.02 0.02 & 10.34 0.02 & 10.11 0.04 & 10.51 0.02 & 10.58 0.02 \\ 
UGC 08204 & 0.023853 & 2017hn & Ia & 4.66 & 0.091 & 17.35 0.04 & 16.03 0.01 & 14.54 0.00 & 13.79 0.00 & 13.38 0.00 & 13.06 0.00 & 11.90 0.04 & 11.17 0.05 & 10.86 0.08 & 11.26 0.03 & 11.15 0.02 \\ 
ESO 440-G001 & 0.028717 & ATLAS17ajn & Ia & 18.48 & 0.206 & 17.40 0.05 & --- & --- & --- & --- & --- & 13.29 0.06 & 12.57 0.07 & 12.36 0.11 & 12.79 0.03 & 12.85 0.03 \\ 
2MASX J08150520+3811205 & 0.054000 & MASTER OT J081506 & Ia & 11.35 & 0.101 & 18.94 0.06 & 16.95 0.03 & 15.46 0.01 & 14.77 0.00 & 14.28 0.00 & 13.99 0.01 & 13.80 0.06 & 13.16 0.09 & 12.66 0.09 & 13.32 0.04 & 13.10 0.04 \\ 
2MASX J00573150+3011098 & 0.016331 & ATLAS17air & Ia & 5.88 & 0.201 & 19.21 0.08 & 18.20 0.04 & 16.72 0.00 & 16.03 0.00 & 15.71 0.00 & 15.47 0.01 & 14.74 0.09 & 14.19 0.14 & 13.88 0.14 & 14.06 0.03 & 14.04 0.04 \\ 
NGC 5541 & 0.025678 & 2017mf & Ia & 12.6 & 0.030 & 15.84 0.02 & --- & --- & --- & --- & --- & 11.17 0.01 & 10.43 0.02 & 10.07 0.02 & 11.00 0.03 & 10.82 0.02 \\ 
SDSS J102641.99+364053.2 & 0.024639 & PTSS-17dfc & Ia & 5.22 & 0.027 & 19.01 0.06 & 18.23 0.03 & 17.09 0.01 & 16.67 0.01 & 16.44 0.01 & 16.34 0.03 & $>$17.0 & $>$16.4 & 14.59 0.06* & 15.10 0.04 & 15.13 0.08 \\ 
SSTSL2 J235328.89+034400 & 0.038800 & ATLAS17akw & Ia-91T & 34.38 & 0.136 & --- & 21.96 0.27 & 20.42 0.03 & 19.24 0.02 & 18.69 0.02 & 18.32 0.04 & $>$17.0 & $>$16.4 & 14.48 0.06* & 14.99 0.04 & 14.30 0.05 \\ 
2MASX J02491020+1436036 & 0.027900 & ATLAS17alb & Ia & 2.88 & 0.299 & 19.51 0.11 & --- & --- & --- & --- & --- & 13.31 0.04 & 12.57 0.04 & 12.23 0.07 & 12.29 0.03 & 12.24 0.03 \\ 
Uncatalogued & 0.014000 & ATLAS17amz & IIP & 0 & 0.384 & --- & --- & --- & --- & --- & --- & $>$17.0 & $>$16.4 & 14.84 0.05* & 15.35 0.03 & 15.19 0.07 \\ 
2MASX J13324217-2148034 & 0.02947 & ATLAS17auc & Ia & 1.38 & 0.232 & 19.34 0.10 & --- & --- & --- & --- & --- & 13.66 0.05 & 12.90 0.05 & 12.56 0.09 & 13.06 0.04 & 13.05 0.04 \\ 
GALEXASC J134322.97 & 0.030000 & ATLAS17axb & Ia & 3.84 & 0.300 & 20.96 0.33 & --- & --- & --- & --- & --- & $>$17.0 & $>$16.4 & 13.61 0.05* & 14.12 0.03 & 13.98 0.04 \\ 
KUG 0945+674 & 0.01305 & Gaia17aiq & IIb & 38.6 & 0.327 & --- & --- & --- & --- & --- & --- & $>$17.0 & $>$16.4 & 14.67 0.05* & 15.18 0.03 & 14.91 0.06 \\ 
NGC 3318 & 0.009255 & DLT17h & II & 40.32 & 0.212 & --- & --- & --- & --- & --- & --- & 10.07 0.02 & 9.41 0.03 & 9.10 0.03 & 10.35 0.02 & 10.23 0.02 \\ 
2MASX J08325728-0351295 & 0.030584 & MASTER OT J083256 & Ia & 5.82 & 0.100 & 20.41 0.14 & 17.32 0.02 & 15.47 0.00 & 14.64 0.00 & 14.25 0.00 & 13.96 0.00 & 12.70 0.04 & 12.12 0.05 & 11.82 0.09 & 12.11 0.02 & 12.17 0.02 \\ 
CGCG 270-047 & 0.029037 & PS17bbn & Ia-91bg & 19.44 & 0.052 & 18.50 0.06 & 16.41 0.01 & 14.60 0.00 & 13.83 0.00 & 13.41 0.00 & 13.11 0.00 & 12.34 0.03 & 11.64 0.04 & 11.34 0.06 & 11.67 0.02 & 11.67 0.02 \\ 
CGCG 308-036 & 0.016000 & iPTF17aub & II & 10.02 & 0.229 & 17.59 0.03 & 16.44 0.02 & 15.11 0.00 & 14.49 0.00 & 14.18 0.00 & 13.98 0.01 & 13.33 0.05 & 12.73 0.08 & 12.48 0.10 & 13.21 0.03 & 13.13 0.03 \\ 
CGCG 421-034 & 0.027426 & ATLAS17bam & Ia & 23.46 & 0.319 & --- & 16.61 0.01 & 14.69 0.00 & 13.76 0.00 & 13.32 0.00 & 12.99 0.00 & 11.86 0.03 & 11.18 0.03 & 10.84 0.06 & 11.17 0.02 & 11.19 0.02 \\ 
\vspace{-0.22cm}
 & & & & & & & & & & & & & & \\
\hline
\end{tabular}
\smallskip
\\
\raggedright
\noindent This table is available in its entirety in a machine-readable form in the online journal. A portion is shown here for guidance regarding its form and content. Uncertainty is given for all magnitudes, and in some cases is zero. ``MASTER'' supernova names and ``GALEXASC'' galaxy names have been abbreviated.\\
$^a$ Galactic extinction taken from \citet{schlafly11}. \\
$^b$ No magnitude is listed for those galaxies not detected in GALEX survey data. \\
$^c$ No magnitude is listed for those galaxies not detected in SDSS data or those located outside of the SDSS footprint. \\
$^d$ For those galaxies not detected in 2MASS data, we assume an upper limit of the faintest galaxy detected in each band from our sample. \\
$^e$ $K_S$-band magnitudes marked with a ``*'' indicate those estimated from the WISE $W1$-band data, as described in the text. \\
\end{minipage}
\vspace{-0.5cm}
\end{table}

\end{landscape}

\end{document}